\pdfoutput=1
\documentclass[fleqn,usenatbib]{mnras}

\usepackage{newtxtext,newtxmath}
\usepackage[T1]{fontenc}
\usepackage{ae,aecompl}
\usepackage{multirow}

\usepackage{graphicx}	
\usepackage{amsmath}	
\usepackage{hyperref}   

\newcommand{\avir}{\alpha_{\rm vir}}
\newcommand{\msun}{M$_\odot$}
\newcommand{\mistix}{\texttt{MIStIX}}
\newcommand{\amuse}{\texttt{AMUSE}}
\newcommand{\kms}{km\,s$^{-1}$}
\newcommand{\rvir}{R_{\rm vir}}
\newcommand{\mcluster}{M_{\rm c}}

\title[Binary Star Cluster Formation]{The Formation of Binary Star Clusters in The Milky Way and Large Magellanic Cloud}

\author[Darma, Arifyanto \& Kouwenhoven]{
R. Darma,$^{1}$\thanks{E-mail: darmarendy@gmail.com}
M.I. Arifyanto,$^{2}$\thanks{E-mail: ikbal@as.itb.ac.id}
and M.B.N. Kouwenhoven$^{3}${\thanks{E-mail: t.kouwenhoven@xjtlu.edu.cn}}
\\
$^{1}$Department of Astronomy, Faculty of Mathematics and Natural Sciences, Institut Teknologi Bandung, Jl. Ganesha no. 10, Bandung 40132, Indonesia\\
$^{2}$Astronomy Research Division, Faculty of Mathematics and Natural Sciences, Institut Teknologi Bandung, Jl. Ganesha no. 10, Bandung 40132, Indonesia\\
$^{3}$Department of Physics, School of Science, Xi'an Jiaotong-Liverpool University, 111 Ren'ai Rd, Suzhou Industrial Park, Suzhou 215123, P.R. China\\
}

\date{Accepted --. Received --; in original form --}
\pubyear{2020}

\begin{document}
\label{firstpage}
\pagerange{\pageref{firstpage}--\pageref{lastpage}}
\maketitle

\begin{abstract}
Recent observations of young embedded clumpy clusters and statistical identifications of binary star clusters have provided new insights into the formation process and subsequent dynamical evolution of star clusters. The early dynamical evolution of clumpy stellar structures provides the conditions for the origin of binary star clusters. Here, we carry out $N$-body simulations in order to investigate the formation of binary star clusters in the Milky Way and in the Large Magellanic Cloud (LMC). We find that binary star clusters can form from stellar aggregates with a variety of initial conditions. For a given initial virial ratio, a higher degree of initial substructure results in a higher fraction of binary star clusters. The number of binary star clusters decreases over time due to merging or dissolution of the binary system. Typically, $\sim 45\%$ of the aggregates evolve into binary/multiple clusters within $t=20$~Myr in the Milky Way environment, while merely $\sim30\%$ survives beyond $t=50$~Myr, with separations $\lesssim 50$~pc. On the other hand, in the LMC, $\sim 90\%$ of the binary/multiple clusters survive beyond $t=20$~Myr and the fraction decreases to $\sim 80\%$ at $t=50$~Myr, with separations $\lesssim 35~$pc. Multiple clusters are also rapidly formed for highly-substructured and expanding clusters. The additional components tend to detach and the remaining binary star cluster merges. The merging process can produce fast rotating star clusters with mostly flat rotation curves that speed up in the outskirts.
\end{abstract}



\section{Introduction}
\label{sec:1}

Star clusters is among the most important astrophysics laboratories for understanding the formation process of stars, the dynamics of planets and stars, the evolution of single and binary stars, and galactic structure and dynamics. Dense stellar environments are commonly the regions where the stars are born (see \citealt{Lada}). These regions typically dissolve within several million years (\citealt{deGrijsParmentier2007}; \citealt{ParmentierdeGrijs}). As such, star clusters can be considered as the building blocks of galaxies (\citealt{Martinez-Barbosa2016}). 

Numerous observational surveys have been carried out to catalogue the star cluster contents of the Milky Way and of the Magellanic Clouds, such as WEBDA (\citealt{Mermilliod}), the Milky Way Star Clusters Catalogue (MWSC, see \citealt{Kharchenko}), the New Catalogue of Optically Visible Open Clusters and Candidates (NCOVOCC, see \citealt{Dias}), and the catalogue of \cite{Bhatia1991}. Furthermore, previous observations \citep[e.g.,][]{Bhatia, Bhatia1991, Pietrzynski} have identified a substantial number of star cluster pairs that are on nearly the same line-of-sight; these are known as double clusters. The small angular separation between the two components of such pairs often suggests that they gravitationally interact, although there is usually no direct evidence for this, beyond statistical hints. A small number of double clusters were investigated in detail, and were discovered to mutually orbit each other; these double clusters are often referred to as binary star clusters. The most famous binary star cluster in the Milky Way is the {\em Double Cluster}, also known as NGC\,869/NGC\,884 and as $h + \chi$ Persei (\citealt{Messow}; \citealt{Zhong2019}). Other candidates include the pairs NGC 1981/NGC 1976 and NGC 3293/NGC 3324 as suggested by \cite{Marcos2009}. Well-known extragalactic examples of binary star clusters include SL\,538/NGC\,2006 (\citealt{DieballGrebel}), NGC\,1850 (\citealt{Fischer}), and NGC\,2136/NGC\,2137 (\citealt{Hilker}) in the LMC. Recently discovered binary star cluster candidates in the LMC are SL\,349/SL\,353, SL\,385/SL\,387/NGC\,1922, and NGC\,1836/BRHT\,4b/NGC\,1839 (\citealt{Dalessandro}). However, \cite{Dalessandro} also showed that the double cluster SL\,385/SL\,387 with the third component NGC\,1922 and the double cluster NGC\,1836/BRHT\,4b with NGC\,1839 have an angular separation larger than $120$~arcsec. Such a large separation indicates that the influence of external tidal fields has dissociated the constituent star clusters, which have detacheded on a relatively short time scale (\citealt{Sugimoto}; \citealt{Bhatia1990}). More interestingly, \cite{Naufal2020} have identified new three candidates of double cluster in the LMC with separations $\sim 13$~pc, i.e. ASCC 16/ASCC 21, NGC 6716/Collinder 394, and NGC 2547/Pozzo 1 from the second data release of GAIA.

A number of studies have been carried out to statistically detect binary star clusters in the LMC, and to estimate their physical and orbital parameters. \cite{Bhatia} listed 69 double clusters with projected separations $\leq 18$~pc in the LMC. \cite{Pietrzynski} identified the existence of double or multiple clusters in the LMC using the Optical Gravitational Lensing Experiment (OGLE) mission and listed 73 double clusters, 18 triples, and 5 quadruples. In addition, \cite{PietrzynskiUdalski} identified 23 double clusters and 4 triples with projected separations $\leq 20$~pc, and relative young ages, in the LMC, which suggests that these star cluster systems are primordial (\citealt{Dieball}). Moreover, \cite{Dieball} proposed a list of 473 binary star cluster candidates with projected separation $\leq 20$~pc in the LMC that have a large probability of being a component of a double or multiple star cluster. The updated statistics of spatial properties, ages, sizes, and separation distributions from \cite{Priyatikanto2019a} provide evidence that binary star cluster formation induced by cloud-cloud collisions through simultaneous or sequential mechanisms (\citealt{Marcos2009}); they identified 634 binary/multiple star cluster candidates in the LMC. According to \cite{Priyatikanto2019b}, the close encounter of the LMC with the Small Magellanic Cloud (SMC) roughly $150$~Myr ago is one of the factors that has increased the formation rate of binary star clusters in the LMC.

Among the earlier identifications of binary star clusters in the Milky Way is that of \cite{Subramaniam}. This study presents a list of 18 double clusters with separations $\leq 20$~pc, which have small age differences, and similar radial velocities and extinctions. According to \cite{Subramaniam}, a typical giant molecular cloud has a spatial extent of 20~pc, and double clusters with separation smaller than such value have higher probability of being physically associated. From an analysis of the WEBDA and NCOVOCC catalogues, \cite{Marcos2009} concluded that at least 12\% of the open clusters in the solar neighbourhood are currently undergoing a gravitational interaction with another star cluster, which strongly suggest that they may be part of a binary star cluster. These binary star clusters have separations $\leq 30$~pc, which is up to three times the tidal radius for a typical open star cluster in the Milky Way disc \citep[see, e.g.,][]{Innanen, BinneyTremaine}. In addition, \cite{Marcos2009} argued that the two components of binary star clusters appear to form both simultaneously and sequentially, but that the components of star cluster systems of higher multiplicity form only sequentially. From the analysis of several updated catalogues, \cite{Priyatikanto2017} identified 47 clusters   in binary/multiple systems, with projected separation $\leq 30$~pc.

A number of numerical studies have been carried out to investigate the dynamical past and future of binary star clusters. \cite{PortegiesRusli} performed $N$-body simulations to study the past and future evolution of the binary star cluster NGC\,2136/NGC\,2137, which is located in the LMC. In order to account for observational uncertainties, their study included a range of possible initial masses, sizes, and projected separations of the cluster pair. They found that the initial separation of the cluster pair is $15-20$~pc at the time of formation.  Cluster pairs with smaller initial separations resulted in a merger process within $\lesssim 60$~Myr, due to angular momentum loss from escaping stars. With a wider initial separation, on the other hand, the two components  gradually moved away from each other. The process is affected by mass loss due to stellar escapers and stellar evolution. Moreover, mass loss due to stellar evolution during the early phases of the interaction was found to  trigger an increase in the eccentricity of the binary star cluster orbit. Their models, however, did not include the external tidal perturbations from the host galaxy (the LMC).

Another study was carried out by \cite{Marcos2010}, who used $N$-body simulations to constrain the origin of binary star clusters under the influence of the tidal field of the Milky Way. Their initial conditions included a range of orbital elements and mass ratios for the star cluster pairs. They found that a long-term stability of binary star clusters is rare. Binary star clusters with small separations generally result in mergers, while the pairs with wider separations are rapidly separated under the influence of the Galactic tidal field. \cite{Priyatikanto2016} found that the dynamical evolution of star cluster pairs is determined by the initial spatial orientation and the initial orbital phase, with respect to the Galactic centre. Moreover, they also found that the merger experienced by the cluster pairs typically occurs within $\sim$100~Myr, and can result in rapidly rotating star clusters, and that the properties of the rotation curve also depends on the orientation of the system in with respect to the Galactic tidal field.

All of the aforementioned numerical studies modelled the structure of star clusters using spherically symmetric stellar distributions, such as the \cite{Plummer} model. Observational studies, however, have shown that the star-forming regions and young star clusters (with ages typically younger than a few million years) have clumpy spatial structures \citep[e.g.,][]{Hillenbrand, Marcos2006, Chen, Wang, Bik}. The \mistix{} survey further supports that argument, after observing a large number of embedded young star clusters in the Milky Way (see \citealt{Kuhn}). In other words, instead of unraveling the origin of binary star clusters by modelling spherically symmetric stellar distributions, it is more realistic to model substructured young star clusters. Such substructures can be characterised using the fractal substructure parameter \citep[e.g.,][]{GoodwinWhitworth, DaffernPowell}. Such clumpy initial structures of star clusters were seen to trigger the formation of binary star clusters (see \citealt{Arnold2017}), through the mechanism (see \citealt{Marcos2010}). Such mechanism is thought to form  binary star clusters, as identified by \cite{Dieball}, \cite{Marcos2009}, and \cite{Palma}.

\cite{Arnold2017} provided an explanation for the origin of binary star clusters as a consequence of the star-forming regions being highly substructured. They showed that binary star clusters can be formed from star clusters with a variety of initial structures and initial virial ratios ($\avir$). They also found that the highest fraction of binary star clusters is formed from stellar aggregates with both a high degree of substructure and a high $\avir$. Besides that, the degree of initial substructure, $\avir$, contributes to the stellar velocity distribution and therefore also becomes one of the key-factors that affects the formation process of binary star clusters. The simulations of \cite{Arnold2017} did not include the influence of the external galactic tidal field, nor did they include stellar evolution.

In this work, we perform $N$-body simulations of young star clusters with different degrees of initial substructure and different initial virial ratios, in order to investigate the formation and early evolution of binary star clusters in the Milky Way and in the LMC. Our simulations include the influence of galactic tidal field in which the system resides. We also consider the role of stellar evolution to understand its contribution to the formation and early evolution of binary star clusters. We present the initial conditions, the methodology, and the identification process of binary star clusters in Section~\ref{sec:2}. We present and analyse our results in Section~\ref{sec:3}. Finally, we summarise and discuss our findings in Section~\ref{sec:4}.


\section{Methodology and initial conditions}
\label{sec:2}

\begin{figure*}
    \centering
    \includegraphics[width=0.9\textwidth]{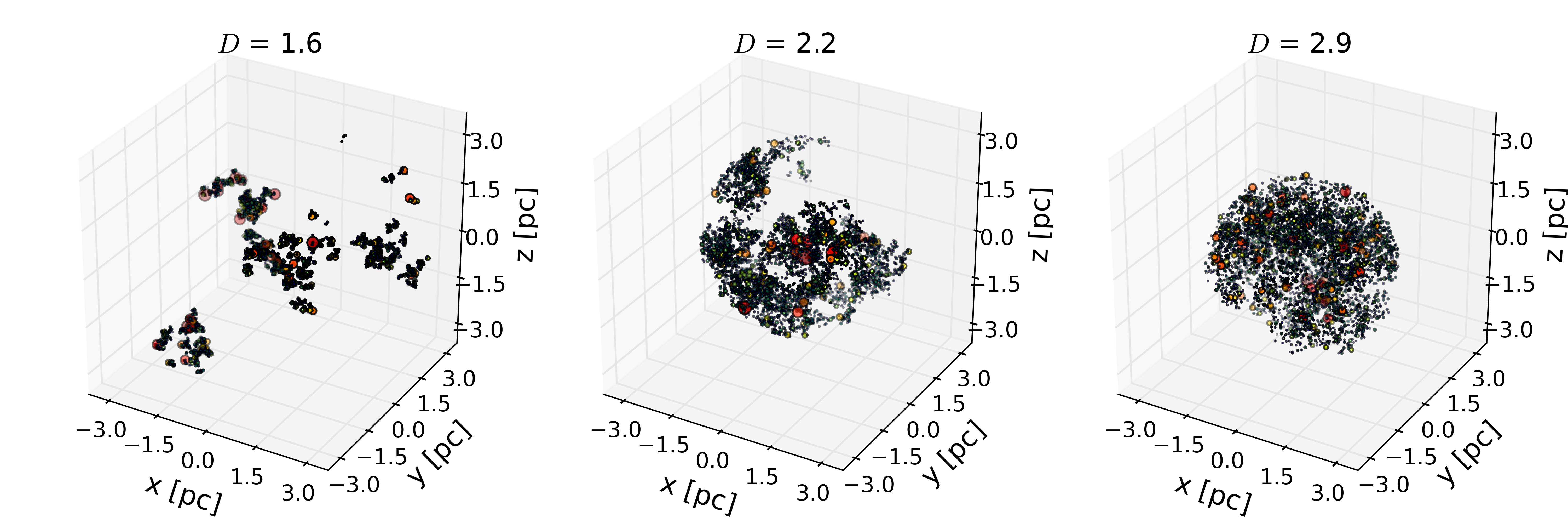}
    \caption{Spatial distributions of the stellar population for three realisations of star clusters with $D = 1.6$ ({\em left panel}), 2.2 ({\em middle panel}), and 2.9 ({\em right panel}) for $\avir$ = 0.5. Each dot represents a star, where the size of each dot scales with the mass of the star.}
    \label{fig:1}
\end{figure*}

\subsection{Star cluster model}

\begin{table}
	\begin{center}
		\caption{Initial conditions for the star clusters (see also Table~\ref{tab:2}).}
	    \label{tab:1}
		\begin{tabular}{l l}
			\hline
			Parameter & Value \\
			\hline
			Number of stars & $N=5000$ \\
			Initial virial radius & $\rvir=2$~pc \\
			Spatial distribution & Substructured (see Table~\ref{tab:2}) \\
			Fractal dimension & $1.6\leq D \leq 2.9$ (see Table \ref{tab:2}) \\
			Virial ratio & $0.3\leq \avir \leq 0.7$ (see Table~\ref{tab:2}) \\
			Initial total mass  & $\mcluster= 2651-3626$~\msun \\
			Initial mass function & \cite{Kroupa} IMF; $0.1-50$~\msun \\
			External tidal field & Isolated; Milky Way; LMC \\
			\hline
		\end{tabular}
	\end{center}
\end{table}

\begin{table}
	\begin{center}
		\caption{Initial conditions of star cluster set from the combinations of $\avir$ and $D$. Each condition is denoted as C1$-$C9 (see also Table~\ref{tab:1}).}
		\label{tab:2}
		\begin{tabular}{c c c c}
			\hline
			Model & $\avir$ & $D$ \\
			\hline
			C1 & 0.3 & 1.6 \\
			C2 & 0.3 & 2.2 \\
			C3 & 0.3 & 2.9 \\
			C4 & 0.5 & 1.6 \\
			C5 & 0.5 & 2.2 \\
			C6 & 0.5 & 2.9 \\
			C7 & 0.7 & 1.6 \\
			C8 & 0.7 & 2.2 \\
			C9 & 0.7 & 2.9 \\
			\hline
		\end{tabular}
	\end{center}
\end{table}

The virial radii of typical young open clusters is of the order $\rvir \approx 1$~pc (corresponds to a half-mass radius of $R_{\rm hm}\approx 0.77$~pc) and the typical cluster total mass is $\mcluster \approx 1500$~\msun{}  \cite[see, e.g.,][]{HeggieHut2003}. There is, however, a large spread in the sizes of observed star clusters \citep[e.g.,][]{Lada, Schilbach2006, Portegies2010}. In our study, we model slightly larger and more massive star clusters. The initial conditions are listed in Table \ref{tab:1}. 
For all our models, we adopt an initial virial radius of $\rvir = 2$~pc 
($R_{\rm hm} \approx 1.54$~pc), and we initialise each model with $N=5000$ stars. The \cite{Kroupa} initial mass function is implemented to generate the cluster stellar masses with a lower cut-off mass of 0.1~\msun\, and a maximum stellar mass of 50~\msun. The typical average mass of the modelled star clusters is $\big \langle \mcluster \big \rangle \approx 3051$~\msun. Our models are somewhat more massive than those of previous numerical studies (\citealt{GoodwinWhitworth}; \citealt{Yu2011}; \citealt{Arnold2017}). We adopt a Solar metallicity of $Z$ = 0.02 \citep[see, e.g.,][]{Martinez-Barbosa2016}.

We follow the prescriptions of \cite{GoodwinWhitworth} to initialise the positions in each substructured star cluster. In order to generate the fractal stellar distribution, we define a cube of length $N_{\rm div}$ and place one particle as the first-generation parent-particle at the centre of the cube. The cube is then divided into $N_{\rm div}^3$ subcubes. A child-particle is placed at the centre of each subcube, and the first-generation parent-particle is removed. At the next step, we take each child particle which has a probability corresponding to ${N_{\rm div}}^{D-3}$ as the second-generation parent-particle, while others are deleted from the cube. Then a little noise is added to each particle's position, to obtain a non-gridded structure. For the next-generations of parent particles, this process is repeated recursively, until a sufficiently large number of particles is obtained. The particles are then pruned and randomly removed in order to produce a sphere from the cube, until the required number of particles is left. The surviving particles are identified with the stars of the cluster.

The quantity $D$ is the fractal dimension parameter, which describes the degree of substructure of star cluster, and can take values in the interval $D\in[0,3]$. Smaller values indicate a clumpier structure of the star cluster. We also adopt the procedure of \cite{GoodwinWhitworth} to initialise the stellar velocities. The stellar velocity distribution within star cluster is more likely coherent. We summarise the physical parameters used to initialise the star cluster models in Table \ref{tab:1}.

We study star clusters with different initial virial ratios ($\avir$) and initial degrees of substructure ($D$), as in \cite{Arnold2017}; see Table~\ref{tab:2}. Here, $\avir$ is the initial virial ratio of the star cluster (i.e., the ratio of the total kinetic to the total potential energy). Higher values of $\avir$ correspond to higher stellar velocity dispersions. Examples of the spatial distributions of the stars in models with different conditions are illustrated in Figure~\ref{fig:1}.

We study the formation process of binary stars cluster in three different environments: (i) isolated star clusters, (ii) star clusters in the tidal field of the Milky Way, and (iii) star clusters in the tidal field of the LMC (see Section~\ref{section:tidalmodels} for details). For each scenario, we explore a range of initial conditions for the star clusters, as listed in Table~\ref{tab:2}. We adopt identical initial stellar positions, stellar velocities, and stellar masses for each of the three scenarios. 
For simplicity, we assume that all primordial gas has been expelled from the star clusters, that there are no primordial binaries, and that no primordial mass segregation is present. We carry out simulations of an ensemble of thirty realisations for each set of initial conditions. 


\subsection{Milky Way and LMC models} \label{section:tidalmodels}

\begin{table}
	\begin{center}
		\caption{The parameters of Milky Way potential, adopted from \protect\cite{Haghi}.}
		\label{tab:3}
		\begin{tabular*}{\linewidth}{@{\extracolsep{\fill}}p{0.3\linewidth}p{0.3\linewidth}p{0.3\linewidth}@{}}
			\hline
			Bulge & Disc & Dark Matter Halo \\
			\hline
			$M_{\rm B} = 2.5 \times 10^{10}$ \msun & $M_{\rm D} = 7.5 \times 10^{10}$ \msun & $M_{\rm vir} = 9 \times 10^{11}$ \msun \\
			$r_{\rm c} = 0.5$ kpc & $a = 5.4$ kpc & $\rvir = 250$ kpc \\
			& $b = 0.3$ kpc & $c = 13.1$ \\
			\hline
		\end{tabular*}
	\end{center}
\end{table}

\begin{table}
	\begin{center}
		\caption{The parameters of LMC potential, adopted from \protect\cite{Salem}.}
		\label{tab:4}
		\begin{tabular}{l l}
			\hline
			Disc & Dark Matter Halo \\
			\hline
			$M_{\rm D} = 2.7 \times 10^9$ \msun & $\rho_{\rm 0} = 3.4 \times 10^{-24}$ g\,cm$^{-3}$ \\
			$a = 1.7$ kpc & $r_{\rm 0} = 3$ kpc \\
			$b = 0.34$ kpc & \\
			\hline
		\end{tabular}
	\end{center}
\end{table}

We model the Milky Way potential following the prescriptions of \cite{Haghi}. The Milky Way potential consists of a bulge, a disc, and a dark matter halo which we adopt from \cite{Hernquist}, \cite{MiyamotoNagai}, and \cite{Navarro}, respectively. The gravitational potential of the bulge is modelled as
\begin{equation} \label{eq:1}
	\Phi_{\rm B} = -\frac{GM_{\rm B}}{r + r_{\rm c}} \quad ,
\end{equation}
the gravitational potential of the disc as 
\begin{equation} \label{eq:2}
	\Phi_{\rm D} = -\frac{GM_{\rm D}}{\sqrt{x^2 + y^2 + \left ( a + \sqrt{z^2 + b^2} \right )^2}} \quad ,
\end{equation}
and the gravitational potential of the dark matter halo of the Milky Way is modelled as
\begin{equation} \label{eq:3}
	\Phi_{\rm H}(r) = -\frac{GM_{\rm vir}}{r \left [ \log{(1+c)} - \frac{c}{1+c} \right ]} \log{\left ( 1+\frac{cr}{\rvir} \right )} \quad .
\end{equation}
In these expressions, $r = (x^2 + y^2 + z^2)^{1/2}$ is distance to the galaxy centre, which is located at the origin. The masses of the bulge, disc, and dark matter halo are denoted as $M_{\rm B}$, $M_{\rm D},$ and $M_{\rm vir}$, respectively. Furthermore, $a$ is the scale length of the disc, and $b$ is the scale height of the disc. The scale radius of the bulge and the virial radius of the dark matter halo are denoted as $r_{\rm c}$ and $\rvir$, respectively. The parameters $c$ and $G$ are the scale radius of the dark matter halo and the gravitational constant, respectively. The values of all parameters of the Milky Way potential were obtained from \cite{Haghi}, and are listed in Table~\ref{tab:3}. In the Milky Way environment, we adopt a Solar Galactic orbit, i.e. with initial position $\vec{r}$ = ($-8.5, 0.0, 0.02$)~kpc (\citealt{Martinez-Barbosa2017}) and velocity $\vec{v}$ = ($11.2, 12.4 + V_{\rm LSR}, 7.25$)~\kms{} (\citealt{Schonrich}). Here, $V_{\rm LSR} = 220$~\kms{} is the velocity of Local Standard of Rest (LSR) which the value is corresponding to the Milky Way potential model. 

The gravitational potential of the LMC is modelled following the prescriptions of \cite{Salem}, and consists of a disc and a dark matter halo. The disc and dark matter halo are modelled using the prescriptions of \cite{MiyamotoNagai} as seen in Eq.~(\ref{eq:2}) and of \cite{Burkert} as seen in Eq.~(\ref{eq:4}). In fact that \cite{Salem} also modelled the gas potential of LMC, but as mentioned above, we do not consider the presence of gas in this work. We model the mass density profile of the LMC, $\rho_{\rm H}(r)$ as
\begin{equation} \label{eq:4}
	\rho_{\rm H}(r) = 
	\frac{
	\rho_{\rm 0} {r_{\rm 0}}^3 
	}{
	(r + r_{\rm 0}) (r^2 + {r_{\rm 0}}^2)
	} \quad .
\end{equation}
Here, $\rho_{\rm H}$ is the mass density of the dark matter halo, $\rho_{\rm 0}$ is the corresponding density at the centre of the LMC, and $r_{\rm 0}$ is the scale radius of the dark matter halo.  The expression above is transformed into a potential function using the Poisson equation, i.e. $\nabla^2 \Phi = 4 \pi G \rho$ (see, e.g., \citealt{BinneyTremaine}). We adopt the parameters of LMC potential from \cite{Salem} (see Table~\ref{tab:4}).
For star clusters in the LMC, we place the star clusters at roughly half  the radius of the LMC ($\sim$4.3 kpc, see, e.g., \citealt{Vaucouleurs}), i.e. we assign an initial position $\vec{r}$ = ($-2.0, 0.0, 0.0$)~kpc and velocity $\vec{v}$ = ($0.0, -70.0, 0.0$)~\kms{}. The latter values were extracted from the LMC rotation curve \cite{Vasiliev} using the second data release of GAIA.

Under the influence of the galactic tidal field, for a star cluster with a mass $\mcluster$ that orbiting a galactic centre in a circular orbit at a galactocentric distance $R_{\rm g}$, the tidal radius ($r_{\rm t}$) of the star cluster can be calculated with the Jacobi radius:
\begin{equation} 
	r_{\rm t} = R_{\rm g} \left ( \frac{\mcluster}{3M_{\rm g}} \right )^\frac{1}{3}
\end{equation}
(see, e.g., \citealt{Portegies1999}; \citealt{BinneyTremaine}). Here, we adopt $M_{\rm g} = 1.75 \times 10^{10}$~\msun\,as the mass of the Milky Way at $R_{\rm g} = 8.5$~kpc and $M_{\rm g} = 6.99 \times 10^8$~\msun\,as the mass of the LMC at $R_{\rm g} = 2.0$~kpc. For a star cluster with an average initial total mass of $\big \langle \mcluster \big \rangle \approx 3051$~\msun, the initial tidal radius typically is $r_{\rm t} \approx 33$~pc in the Milky Way environment, and $r_{\rm t} \approx 23$~pc in the LMC environment.


\subsection{Numerical method} \label{subsec:1}

The $N$-body simulations are performed using the \amuse{} code\footnote{Detailed instructions for modelling various astrophysics processes using \amuse{}, and the implementations of codes, can be found at \href{https://amusecode.github.io/}{https://amusecode.github.io/}.} (see, \citealt{PortegiesZwart}; \citealt{Pelupessy}; \citealt{Sills}; \citealt{Helm}; \citealt{Cai}; \citealt{Darma2019a}, \citealt{Darma2019b}). The gravitational interactions between stars are calculated using the \texttt{BHTree} code (\citealt{BarnesHutt}). We implement a softening length, $\epsilon = R_{\rm vir}N^{-1/3}$, where $N$ is the initial number of stars in the model. The \texttt{BHTree} code is coupled with the galactic tidal field using the \texttt{Bridge} code (\citealt{Fujii}). Stellar evolution is modelled using the single-star evolution (\texttt{SSE}) code (\citealt{Pols}; \citealt{Hurley}). The majority of the binary star clusters identified by \cite{Marcos2009} have ages younger than 25~Myr. In order to investigate the dynamical fate of binary star clusters at older ages, we therefore perform our simulations for a time span of 50 Myr, with an output interval of 0.1~Myr. 

\begin{figure*}
    \centering
    \includegraphics[width=0.85\textwidth]{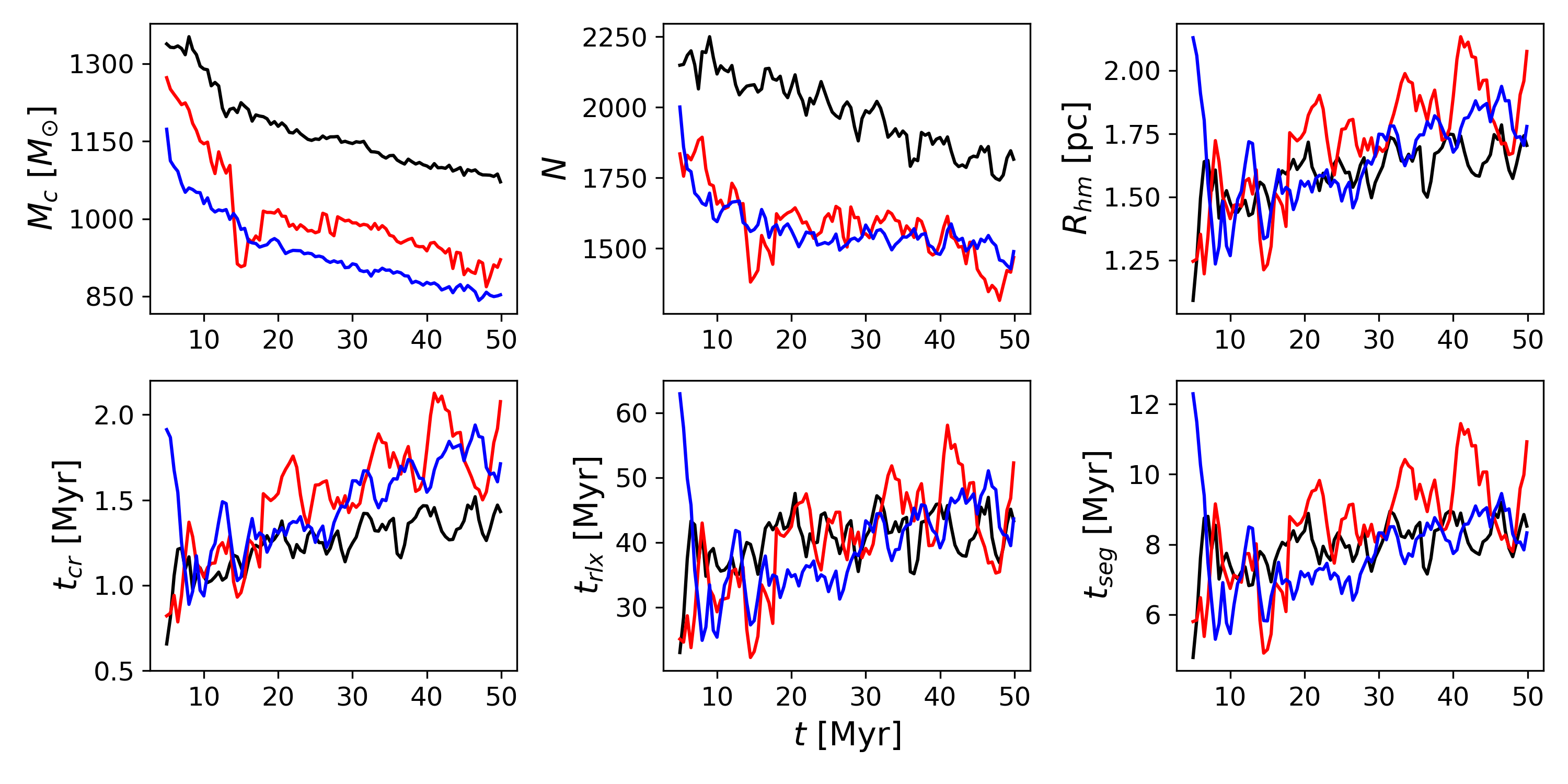}
    \caption{The evolution of the global properties of star clusters in the absence of an external tidal field. The black, red, and blue curves show the evolution for representative examples in the ensemble of realisations for models C1, C4, and C7, respectively. Here, $R_{\rm hm}$ is the half-mass radius of the star cluster, and $N$ and $M_c$ are the number of stars and mass included within radius of $R = 2~$pc from the cluster centre, respectively. The timescales shown in this figure are calculated using the entire stellar population.}
    \label{fig:2}
\end{figure*}


\subsection{Identification of Binary Star Clusters}

We implement the Minimum Spanning Tree (MST) algorithm (\citealt{Prim}) to identify the binary star clusters. After obtaining the MST for each stellar grouping, two additional steps i.e., separation and elimination, are required to obtain the locations of the possible star cluster(s). The minimum length, $\lambda_{\rm c}$, is adopted to separate groupings, and the minimum number of nodes, $N_{\rm c}$, is used to eliminate groupings that are considered too small for our study; see \cite{Barrow1985}, \cite{Campana2008}, \cite{Allison2009b}, and \cite{Campana2013} for further details. Here, we set $0 < \lambda_{\rm c} \leq 4.0$ and $N_{\rm c}=100$ for the different initial conditions of the star clusters, and for the different environments in which the star clusters are embedded. The specific value of $\lambda_{\rm c}$ depends on the stellar distribution of each star cluster. Generally, the MST algorithm can identify the clustering through various parameters such as the position, the velocity, the gravitational force, etc. In this work, we adopt the positions of the stars in the MST algorithm, because of this parameter provides the most robust results for our simulations.

In order to obtain more accurate star cluster identifications, we also confirm the presence of each detected binary star cluster through inspection by eye, for each model and at each time step. \cite{Arnold2017} identified the existence of binary star clusters in their simulations at time 20~Myr, with the argument that binary star clusters are generally short-lived. In this work, we focus particularly on the status of the systems at times $t=20$~Myr and $t=50$~Myr, for comparison with the results of \cite{Arnold2017}, and also to investigate the dynamical fate of the stellar aggregates over longer periods of time.

We classify the status of each system at a given time $t$ into four categories: 
\begin{enumerate}
    \item \textit{Single clusters}. A star cluster that is single at time $t$ and has been single for times $<t$.
    \item \textit{Binary/multiple clusters}. The term \textit{binary/multiple} refers to a system that consists of two or more star clusters that are gravitationally interacting at time $t$.
    \item \textit{Separated systems}. When more than one cluster is present, and the system is not gravitationally interacting at time $t$, the system is categorised as a \textit{separated system}.
    \item \textit{Merger products}. A star cluster that is single at time $t$, but has been binary/multiple at some stage in its past, is classified as a \textit{merger product}.
\end{enumerate}
Finally, we identify for each set of initial conditions, and at each time, the fraction of each category, as well as the corresponding statistical uncertainties.
Note that we do not investigate the micro-clusters; these are not included in our classification scheme. Micro-clusters are gravitationally-bound groups of $< 100$~stars (i.e., $\lesssim$ 5\% of the system's initial total mass). For a detailed discussion on the origin and evolution of micro-clusters, we refer to \cite{Bouvier1971}, \cite{Kirk2014}, and \cite{Arnold2017}.


\section{Results}
\label{sec:3}

\subsection{Star Cluster Evolution}

\begin{figure}
    \centering
    \includegraphics[width=0.45\textwidth]{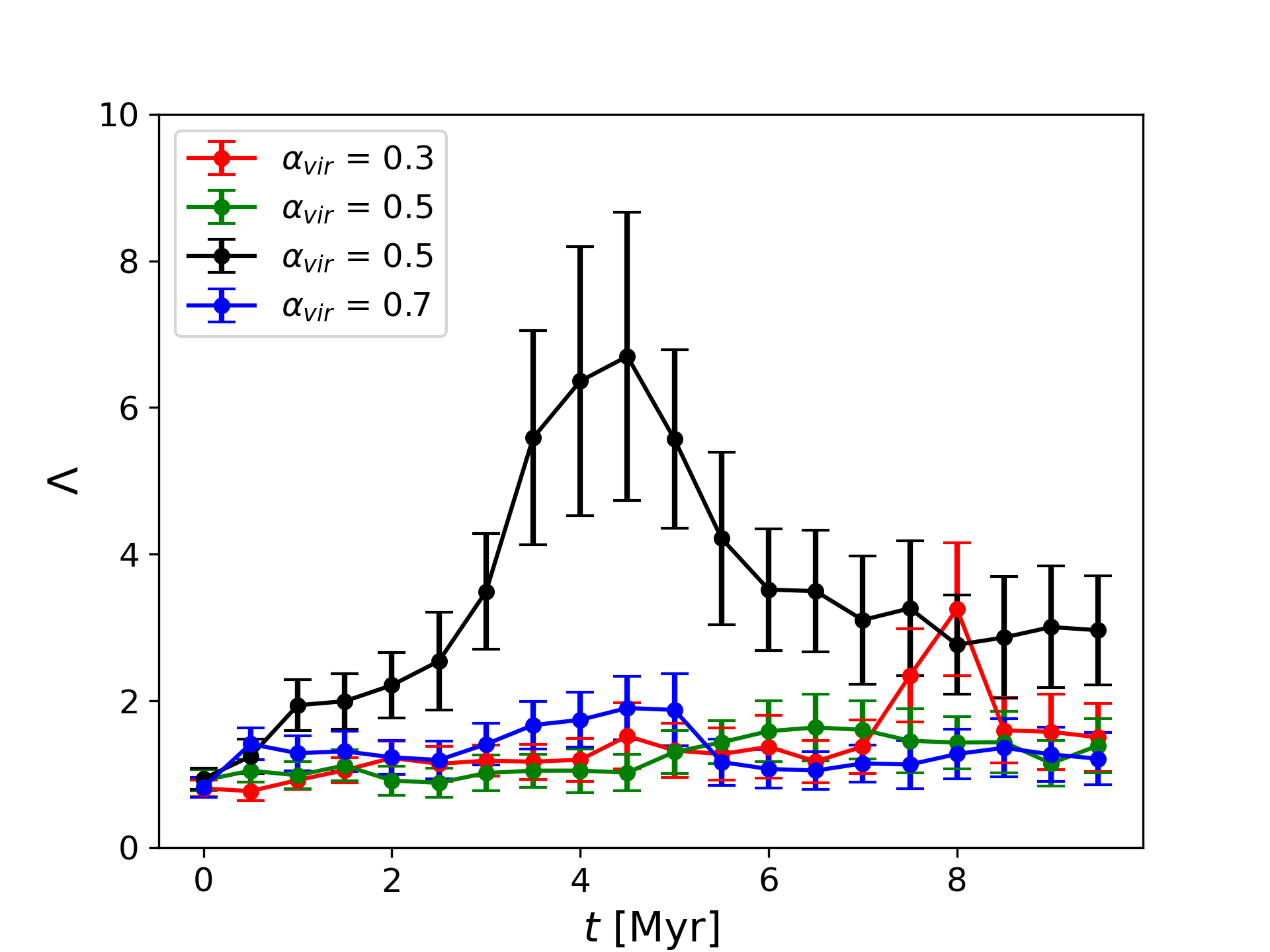}
    \caption{Mass segregation during the first 10~Myr, for star clusters in the isolated environment, with initial conditions C2 (red), C5 (green), and C8 (blue). The black curve represents the result of another realisation of model C5, in which a notably high degree of mass segregation occurs, as a consequence of the merger of stellar agglomerations at early times.}
    \label{fig:3}
\end{figure}

The dynamical evolution of star clusters involves several mechanisms. At early times, the system re-virializes through violent relaxation. The mass loss of star clusters is triggered by the evolution of stellar population and two-body relaxation process, in which also expanses the cluster size (e.g., \citealt{TakahashiPortegies}; \citealt{Baumgardt}; \citealt{Madrid2012}). This expansion phase lasts for roughly 40\% of the star cluster's lifetime (\citealt{Gieles2011}), during which both the crossing time and the relaxation time increase, and the star cluster gradually dissolves.

The crossing time at the half-mass radius in a virialised system, $t_{\rm cr}$, depends on the half-mass radius of the cluster, $R_{\rm hm}$ and the total mass of the cluster, $\mcluster$,
\begin{equation} \label{eq:9}
	t_{\rm cr} = \sqrt{\frac{2}{G} \frac{R_{\rm hm}^3}{\mcluster}} \quad ,
\end{equation}
where $G$ is the gravitational constant. The half-mass relaxation time is given by
\begin{equation} \label{eq:10}
	t_{\rm rlx} \approx \left ( \frac{N}{8 \ln{N}} \right ) t_{\rm cr} 
	\quad ,
\end{equation}
where $N$ is the number of members of a star cluster.

During the first million year, the star cluster experiences violent relaxation and obtains virial equilibrium and a smooth density profile.
Subsequently the star cluster's dynamical evolution is dominated by two-body relaxation following close encounters between stars. These interactions redistribute the energy between the stars and generally result in a migration of the lower-mass stars to the outskirts of the star clusters, where they may escape if they cross the tidal radius. Escape can occur as a result of strong encounters between member stars, or by small perturbations that cause a star to reach the escape velocity and to pass beyond the tidal radius \citep[e.g.,][]{HeggieHut2003}. Over time, the star cluster experiences gradual dissolution, and its member stars become part of the galactic field \citep[e.g.,][]{Gieles2011}. 

The external tidal field experienced by a star cluster strongly affect its dissolution timescale. The rate at which stars escape from the system depends on the cluster's galactocentric distance (\citealt{Madrid2012}), the orbital eccentricity of the cluster's orbit (\citealt{Baumgardt}), the cluster's orbital inclination with respect to the galaxy's plane of rotation (\citealt{Webb}), and the properties of the galactic structure such as the mass and the size of the disc (\citealt{Madrid2014}). 

Two-body relaxation also impacts on the massive stars sinking towards the cluster's centre as a result of a tendency of the stellar population to achieve energy participation (see, e.g., \citealt{Gualandris}; \citealt{Allison2010}). The typical timescale at which a star of mass $M$ experiences mass segregation is
\begin{equation} \label{eq:11}
	t_{\rm seg} \approx \frac{\big \langle m \big \rangle}{M} \frac{N}{8 \ln{N}} t_{\rm cr}
	\quad ,
\end{equation}
where $\big \langle m \big \rangle$ is the average stellar mass \citep[see, e.g.,][]{Spitzer1969}. The average stellar mass for \cite{Kroupa} IMF with a lower stellar mass limit of $M_{\rm cut}$ can be approximated with
\begin{equation} \label{eq:12}
	\big \langle m \big \rangle \approx 0.35 + 2.23 M_{\rm cut} + 0.05 M_{\rm cut}^2 \quad .
\end{equation}
\citep{Kouwenhoven2014}.
Here, for $M_{\rm cut}=0.1$~\msun, we obtain $\big \langle m \big \rangle \approx 0.57$~\msun.
According to \cite{Allison2010}, a substructured star cluster can experience rapid mass segregation due to violent relaxation, resulting in high concentration of $M \approx 2-4$~\msun{} stars in the core. Here, we approximate the mass segregation timescale for $M$ = 3~\msun{} stars in Eq.~(\ref{eq:11}).
The degree of mass segregation in a star cluster can be quantified using a variety of methods. In our study, we express the degree of mass segregation using a parameter $\Lambda$ \citep[see, e.g.,][]{Allison2009b, Allison2009a}, which is commonly defined as
\begin{equation} \label{eq:16}
    \Lambda = \frac{\big \langle l_{\rm norm} \big \rangle}{l_{\rm massive}} \pm \frac{\sigma_{\rm norm}}{l_{\rm massive}}
    \quad .
\end{equation}
Here, $\big \langle l_{\rm norm} \big \rangle$ and $l_{\rm massive}$ are the average lengths of the MSTs of $N$ random stars, and the length of the MSTs of $N$ massive stars that have a mass higher than a chosen limit, respectively. Note that the degree of mass segregation depends on the choice of the mass limit \citep[see, e.g.,][]{Yu2017}. The standard deviation of the MST length is denoted as $\sigma_{\rm norm}$ which is the standard deviation of the MST lengths of the $N$ random stars.

Stellar evolution, internal stellar dynamics, and the interaction of a star cluster with its surroundings ultimately result in the full dissolution of a star cluster. We estimate the longevity of our modelled star clusters using \cite{Lamers2005}'s expression for the dissolution time:
\begin{equation} \label{eq:13}
	t_{\rm diss} \approx 429 \, N^{0.62} \left ( \frac{R_{\rm g}}{V_{\rm g}} \right )
	\  {\rm Myr} \quad .
\end{equation}
In this expression, the galactocentric distance $R_{\rm g}$ is in units of kpc, and the orbital velocity $V_{\rm g}$ in units of \kms.
Our modelled star clusters initially have $\big \langle \mcluster \big \rangle \approx 3051$~\msun\ and $R_{\rm hm} = 1.54$~pc. At time $t=0$~Myr, their dynamical timescales are $t_{\rm cr} = 0.73$~Myr, $t_{\rm rlx} = 53.57$~Myr, and $t_{\rm seg} \approx 7-15$~Myr. Variations in the degree of initial substructure among the modelled clusters  result in star clusters with different measured values for $R_{\rm hm}$ and $t_{\rm cr}$, and hence also affect the range in $t_{\rm seg}$. Note that the values for the mass segregation timescale are rough estimates, as Eq.~(\ref{eq:11}) does not take into account the consequences of initial substructure on the process of mass segregation, which is known to affect the rate at which mass segregation occurs. In the Milky Way environment, the star clusters can typically survive up to $t_{\rm diss} \approx 3.26$~Gyr. For star clusters with identical physical properties, the dissolution time in the LMC environment is somewhat smaller: $t_{\rm diss} \approx 2.41$~Gyr.

We present in Figure~\ref{fig:2} the evolution of properties of single star cluster models with an identical degree of initial substructure, but with different $\avir$. In Figure~\ref{fig:2} shows that the modelled star clusters typically lose $20-30\%$ of their initial total mass during 50~Myr, when evolved in the absence of an external tidal field. This mass loss can be attributed to the loss of typically $15-25\%$ of the member stars through ejection, and to stellar evolution. The mass loss occurs causing the star clusters to expand, and therefore increases the dynamical timescales of the star cluster. Figure~\ref{fig:2} also shows that the variation of $\avir$ does not have significant effect on the evolution of star cluster's dynamical timescales during the 50-Myr timescale covered by the simulations.

\begin{figure*}
    \centering
    \includegraphics[width=0.65\textwidth]{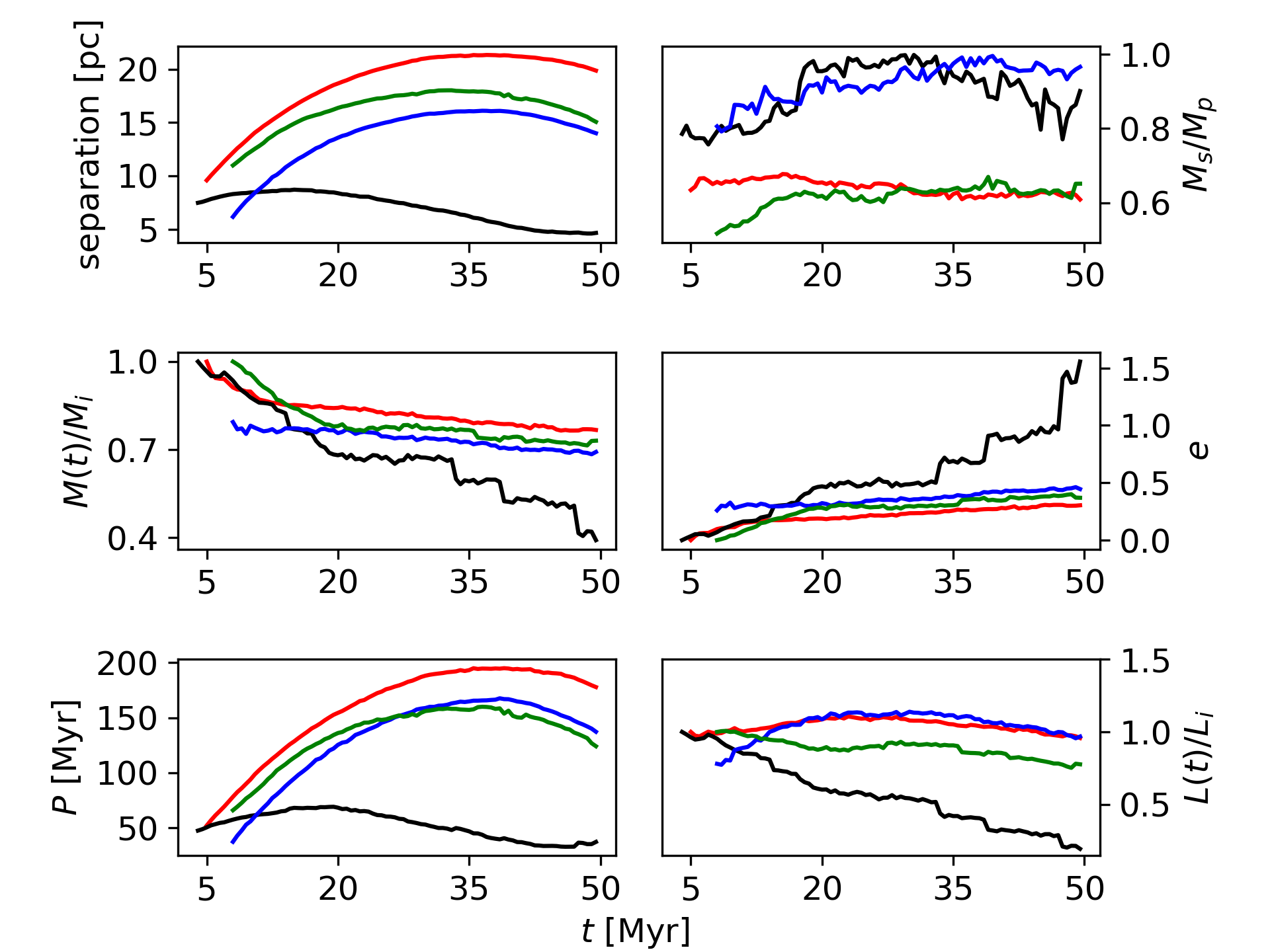}
    \caption{The orbital properties of binary star clusters as a function of time, for the isolated environment, and with initial conditions C1 (red), C4 (green), and C7 (blue). The red, green, and blue curves also illustrate the typical dynamical evolution of binary star cluster found in this work. The black curves represent the result of another realisation of model C4.}
    \label{fig:4}
\end{figure*}

\begin{figure*}
    \centering
    \includegraphics[width=0.65\textwidth]{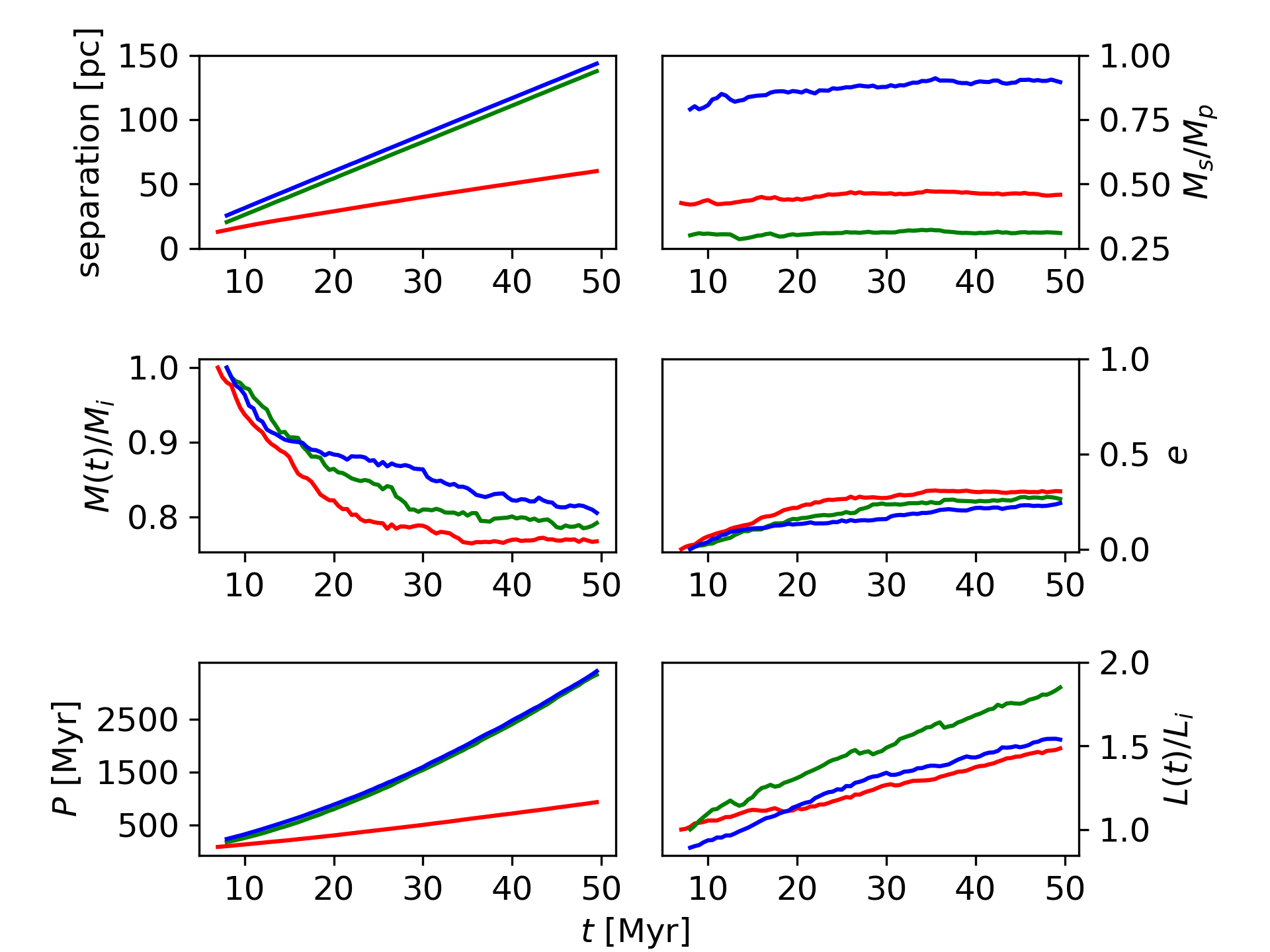}
    \caption{The orbital properties of separated binary star cluster as a function of time. It is the samples for the isolated environment, with initial conditions C1 (red), C4 (green), and C7 (blue).}
    \label{fig:5}
\end{figure*}


We observe a different degree of mass segregation as the clusters evolve during the first 10~Myr; see Figure~\ref{fig:3}. Note that a value $\Lambda=1$ indicates that the cluster has no notable mass segregation. Star clusters with mass segregation have $\Lambda > 1$, while star clusters with inverse mass segregation have $\Lambda < 1$ (see \citealt{Allison2009a}). Several of the modelled star clusters with different degrees of initial substructure form separate agglomerations of stars at early times. These agglomerations then mutually interact through violent relaxation, resulting in a peak in the degree of mass segregation (black curve in Figure \ref{fig:3}). This high degree of mass segregation gradually decreases over time due to stellar evolution. \cite{Yu2011} showed that such behaviour of mass segregation most likely occurs in highly substructured and dynamically cool star clusters at time $t \lesssim 6$~Myr.  Besides that, the rapid merger of binary/multiple clusters and/or the merger of several micro-clusters within a short timescale also enhances the degree of mass segregation, as suggested by \cite{Kirk2014}; see the peak of the red curve in Figure~\ref{fig:3}. However, most of the modelled star clusters do not exhibit a high degree of mass segregation ($\Lambda \approx 1-3$) during the early stages of their evolution. The results indicate that the combinations of $D$ and $\avir$ that form separated stellar agglomerations at early times do not always trigger the violent collapse of these substructures. According to \cite{Yu2011}, the degree of mass segregation at early times is influenced by the properties of the cluster's initial stellar velocity distribution. In several cases, star cluster contains massive stars with a tendency of having a higher velocity dispersion than their lower-mass stars. It can result in an "inverse-mass segregation" \citep[see, e.g.,][]{Allison2009a}, which the expansion of the cluster's outskirts occurs without undergoing a collapse.


\subsection{Evolution of Binary Star Clusters}

Earlier studies of binary star clusters have shown that the separations between the cluster pairs are $\leq 30$~pc in the Milky Way, and are $\leq 20$~pc in the LMC. Separations larger than such values result in separated systems (i.e., two gravitationally unbound single star clusters). Let $M_{\rm p}$ and $M_{\rm s}$ be the mass of the primary component (i.e., the most massive) and the mass of the secondary component, respectively. Let
$M = M_{\rm p} + M_{\rm s}$ be the total mass of the system,  $\mu = M_{\rm p} M_{\rm s} / M$ be the reduced mass of the system, and $q = M_{\rm s}/M_{\rm p}$ be the mass ratio of the system, with $0<q\leq 1$. When the binary star cluster is approximated as two point masses orbiting in a circular orbit, its total orbital angular momentum can be calculated as
\begin{equation} \label{eq:5}
	L = \mu a^2 \omega \equiv \frac{q}{\left ( q + 1 \right ) ^2} a^2 \omega M
\end{equation}
\citep[see, e.g.,][]{PortegiesRusli, Priyatikanto2016}, where $a$ is the semi-major axis and $\omega$ is the orbital angular velocity of the binary system.

As explained by \cite{PortegiesRusli}, mass loss can alter the orbital separation and orbital eccentricity in a two body system. By assuming mass is lost adiabatically and isotropically ($\gamma=1$; see \citealt{PortegiesRusli}), we can determine the system's total mass at any given time, $M(t)$, compared to its initial total mass, $M(0)$:
\begin{equation} \label{eq:6}
	\frac{M(t)}{M(0)} = \left ( \frac{a(t)}{a(0)} \right ) ^{1/(2\gamma - 3)} \quad .
\end{equation}
Here, parameters $a(t)$ is the semi-major axis at time $t$ and $a(0)$ is the initial semi-major axis of the system. Following \cite{Hills1983}, the evolution of the orbital eccentricity at time $t$ due to adiabatic mass loss is
\begin{equation} \label{eq:7}
	e(t) = \frac{M(0)}{M(t)} - 1 \quad ,
\end{equation}
where we have assumed that the orbit is initially circular: $e(0) = 0$.
Note that in our simulations, binary star clusters are formed several million years after the start of the simulations. Thus, such all initial parameters, such as $M(0)$, $a(0)$, and $e(0)$, are calculated at the moment that the binary star cluster is firstly identified.

The changes in masses of the two components and the changes in the semi-major axis during the evolution process of binary star cluster affect the evolution of the orbital period of the system, $P(t)$, over time. For the two star clusters that are orbiting each other in Keplerian orbits, the orbital period of the components can be calculated with an approximation:
\begin{equation} 
	P(t) = 94 \left ( \frac{S(t)}{1 + e(t)} \right ) ^ {\frac{3}{2}} \frac{1}{\sqrt{M_{\rm p}(t) + M_{\rm s}(t)}} 
	\quad {\rm Myr}
\end{equation}
\citep[see][]{Marcos2010}. Here, the unit of mass is \msun{} and the unit of length is pc. The apoclustron is denoted as $S(t)=a(t)\cdot (1+e(t))$.

In Figure~\ref{fig:4} we present the orbital properties of a sample binary star clusters evolving in the absence of an external tidal field. The conditions at times $t< 5$~Myr are more chaotic for star clusters with a higher degree of initial structure. Most of substructure disappears during that period (see, e.g., \citealt{Allison2009a}). We find that all binary star clusters form at early times, immediately after the phase of violent relaxation. After the binary star clusters have formed, the orbital separations generally increases as a consequence of mass loss due to stellar evolution (see \citealt{Sugimoto}) for several million years and subsequently tend to decrease for the remainder of their lifetimes. The same occurs for the orbital periods and orbital angular momenta (red, green, and blue curves in Figure~\ref{fig:4}). Most of the identified binary clusters have maximum separations of less than 30~pc, which is consistent with the findings of previous studies (e.g., \citealt{Subramaniam}; \citealt{Marcos2009}; \citealt{Priyatikanto2017}). In addition, the binary star clusters also experience stellar mass loss, which causes system's total mass to decrease. Depending on the rate at which the mass loss occurs, with respect to the orbital period, this mass loss can also increase the orbital eccentricity of binary star clusters as time passes. Our study shows that most binary star clusters have lifetimes that exceed 50~Myr; longevities of over 100~Myr have been suggested in earlier works (e.g., \citealt{Marcos2009}; \citealt{Priyatikanto2016}; \citealt{Priyatikanto2019a}).

In addition to the common evolutionary pathway of binary star clusters described above, we also find several binary star clusters that orbit each other with gradually decreasing separations (i.e., with in-spiralling orbits) up to times 50~Myr (black curve in Figure~\ref{fig:4}). Such binary star clusters are mostly formed from supervirial stellar aggregates ($\avir = 0.7$), in which the binary system is formed from a chaotic system of multiple sub-clusters. Within a few million years, several sub-clusters then escape from the system, and the remaining binary star cluster will end with a merger (see, e.g., \citealt{Sugimoto}, \citealt{Bhatia1990}). However, this merger process will occur at $t>50$~Myr. \cite{Priyatikanto2016} suggest that, under certain conditions the merger process could even occur beyond $\sim 100$~Myr. These processes trigger the rapid decrease in the orbital angular momentum and in the system's total mass. The orbital eccentricity grows, and even becomes hyperbolic during the final phases of the in-spiral.

\begin{figure*}
    \centering
    \includegraphics[width=0.65\textwidth]{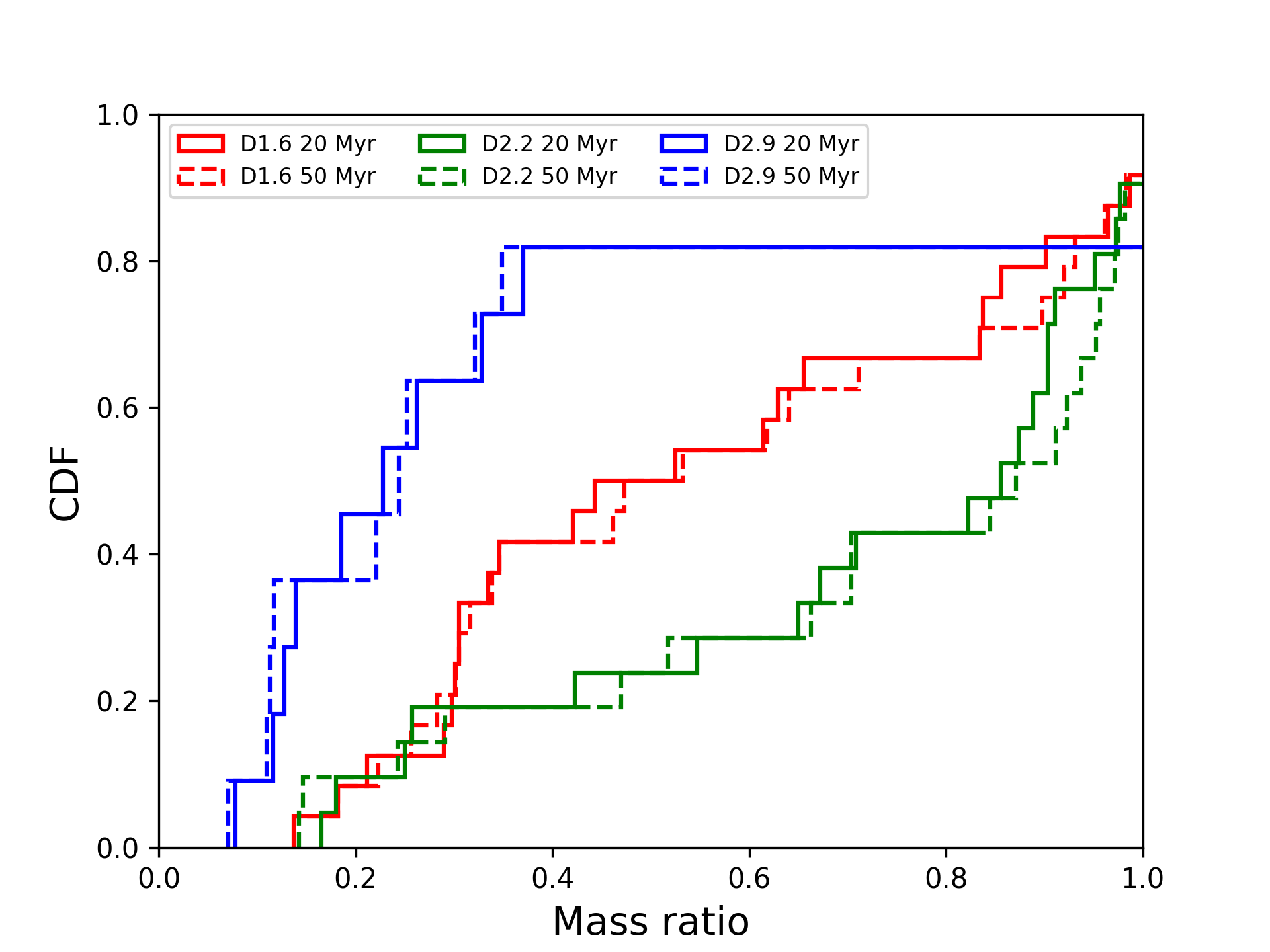}
    \caption{The cumulative distribution of mass ratios for binary star clusters (bound and separated) formed in models with $D = 1.6$ (red), $D = 2.2$ (green), and $D = 2.9$ (blue) at time $t = 20~$Myr (solid histogram) and time $t = 50~$Myr (dashed histogram).}
    \label{fig:6}
\end{figure*}

Several binary star clusters that formed after the chaotic conditions evolved into two distinct, gravitationally unbound star clusters. In Figure~\ref{fig:5} we present the orbital properties of the separated binary star clusters for the isolated environment. The separations between the companions grow (reaching up to 100~pc at $t=50$~Myr) in a prograde orbit which also corresponds to an increasing orbital angular momentum as time passes. In this system, mass loss also occurs, but now it is primarily a consequence of the high orbital angular momentum, rather than stellar evolution. However, mass ratios in both the bound and separated binary star clusters show a high degree of variation. Most clusters that formed from initial substructures with $D=1.6$ have companions with mass ratios $q=0.2-0.4$ with a median $q=0.43$ at $t=20~$Myr. The clusters with initial substructures of $D=2.2$ have mass ratios mostly $q=0.8$ to $q\approx 1$ with a median $q=0.82$ at $t=20~$Myr. For the most clusters with initial substructures of $D=2.9$, the mass ratios are $q=0.1-0.3$ with a median $q=0.19$ at $t=20~$Myr (see Figure~\ref{fig:6}). The mass ratios tend to slightly evolve  between $t=20~$Myr and $t=50~$Myr. Binary clusters with $q \approx 1$ are more often identified from initial substructures with $D=2.2$, while systems with smaller mass ratios are often formed from smaller $D$ and larger $D$. As explained by \cite{Arnold2017}, one may expect to see a sequence in mass ratio distribution shifting from initial substructures of $D = 1.6$, to $D = 2.2$, and to $D = 2.9$. In terms of mergers and binarity, the evolution of structures with larger initial $D$ appears to be dominated by the velocity structure effect, while those with lower initial $D$ appear to be influenced more by the spatial structure of the cluster. These sets of clusters are both likely to form binary clusters with lower mass ratios. With a more optimal correlation between velocity structure and spatial structure in clusters with moderate-$D$, it is more possible to form (roughly) equal-mass binary star clusters.

\subsection{Merger Products}

\begin{figure*}
    \centering
    \includegraphics[width=0.65\textwidth]{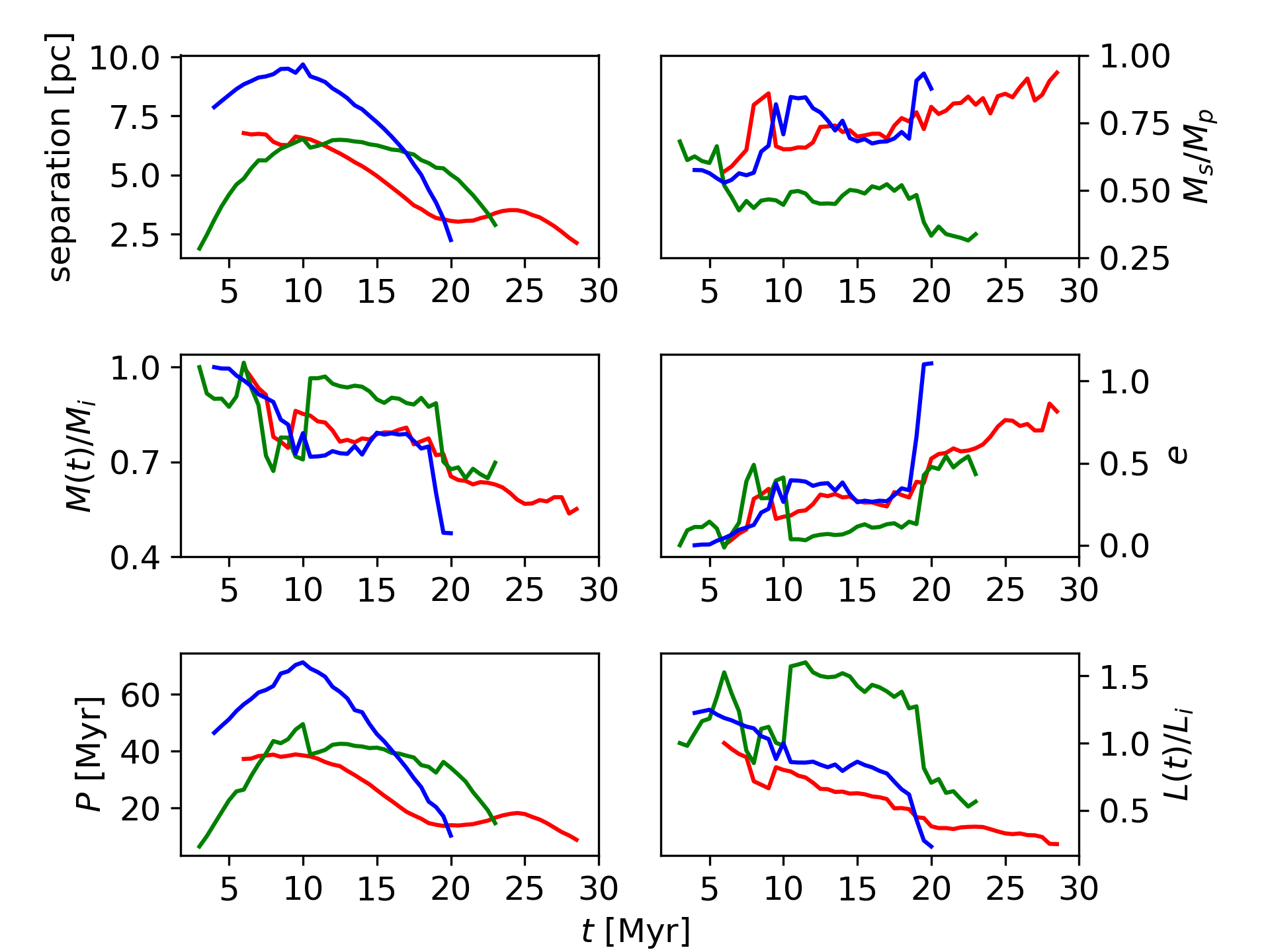}
    \caption{The orbital properties of three binary star clusters in the sample that merge. The red, green, and blue curves represent the dynamical evolution in the isolated environment for initial conditions C1, C4, and C7, respectively.}
    \label{fig:7}
\end{figure*}

A merger between the two components will become more likely when the system's orbital angular momentum decreases in a retrograde orbit (see, e.g., \citealt{Priyatikanto2016}). During the merger process, the cluster orbital angular momentum is mostly transformed into a rotational angular momentum, in which resulting in a rotating merger remnant, as suggested by \cite{Sugimoto} \& \cite{Marcos2010}. In order to understand the rotation of merger remnant, \cite{Priyatikanto2016} analysed the rotation velocities (around the star cluster centre),
\begin{equation} \label{eq:14}
    v_{\rm \theta, i} = \frac{\left ( \vec{r_{\rm i}} \times \vec{v_{\rm i}} \right ) \cdot \hat{L}}{R_{\rm i}} \quad ,
\end{equation}
in the merger remnants, as a function of the distance of the star from the rotation axis,
\begin{equation} \label{eq:15}
    R_{\rm i} = \frac{| \vec{L} \times \vec{r_{\rm i}} |}{| \vec{L} |} \quad .
\end{equation}
Here, $\vec{r_{\rm i}}$ and $\vec{v_{\rm i}}$ are the stars position and stars velocity relative to the cluster's centre of mass. The cluster has a net angular momentum ($\vec{L}$), in which it is the total angular momenta of the cluster's stars ($\vec{L_{\rm i}}$) with masses $m_{\rm i}$, i.e. $\vec{L} = \sum_{i=1}^N (m_i \vec{r_i} \times \vec{v_i})$. If the star cluster has an anisotropy rotational velocity, then $\vec{L}$ will have a non-zero value and the net angular momentum will have direction of $\hat{L} \equiv \vec{L} / |L|$, which denoting cluster's rotation axis.

We present the orbital properties of three merging binary star clusters in Figure~\ref{fig:7}. These binary star clusters merge at times $t=10-35$~Myr as a consequence of a loss of orbital angular momentum. For all mergers that occur in our ensemble of simulations, the mutual separation between the two components of a binary star cluster always remains less than 10~pc. Several of the merger products form during the final phases of in-spiralling orbits, while others form during the collision that occurs at closest approach between the two clusters in a highly eccentricity orbit. The merger process typically takes much longer in in-spiralling orbits compared to eccentric orbits (see, e.g. \citealt{Marcos2010}). Mutual disruption occasionally occurs during the merging process; the disruption is more prominent for merging processes with a longer duration. Instabilities during the merger process cause mass loss of the system. This mass loss tends to be episodic, with  fluctuations that depend on the dynamical behaviour of the cluster and the behaviour of micro clusters surrounding the system.

\begin{figure}
    \centering
    \includegraphics[width=0.45\textwidth]{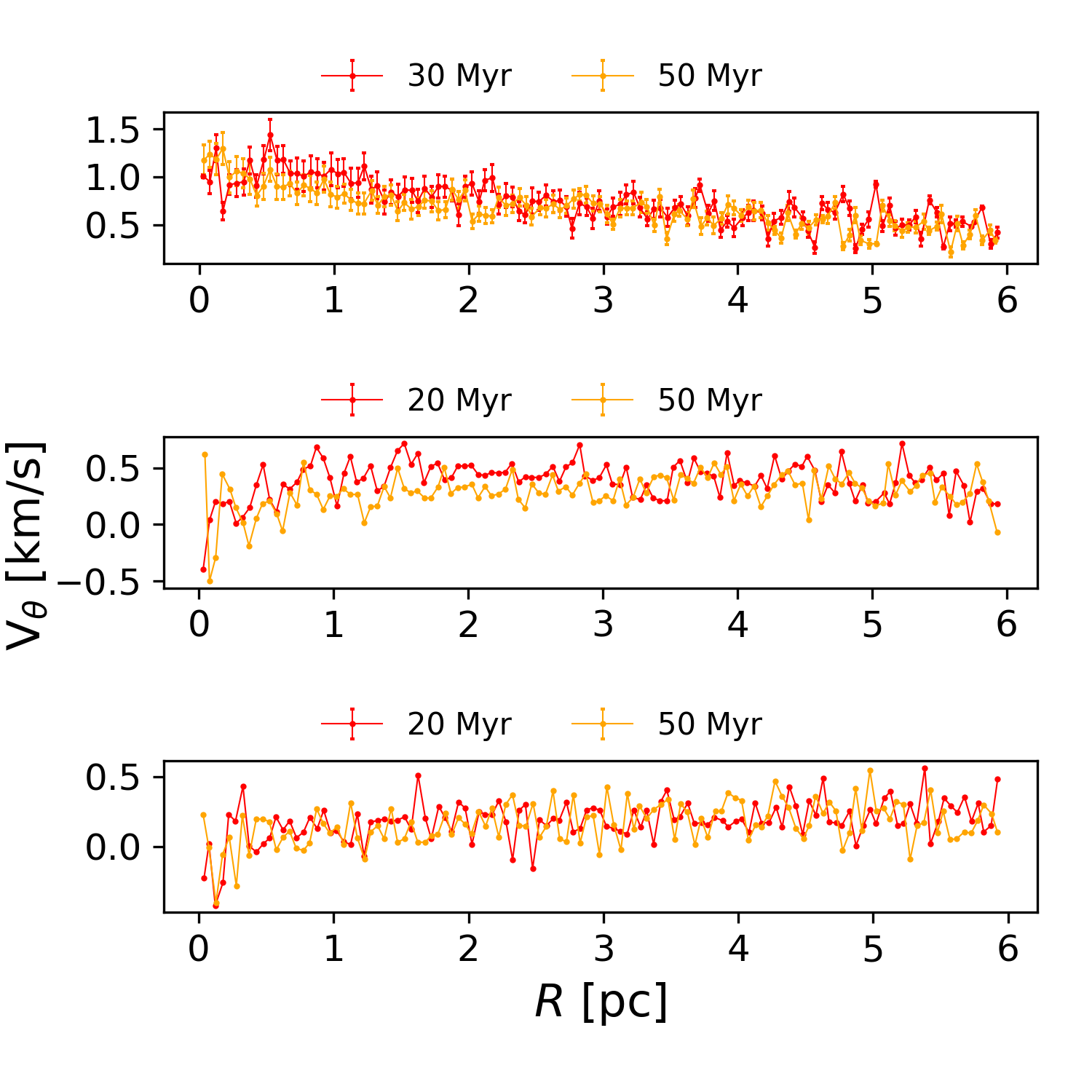}
    \caption{The rotation curves of merger products for the isolated environment with initial conditions C1 (\textit{top panel}), C5 (\textit{middle panel}), and C6 (\textit{bottom} panel).}
    \label{fig:8}
\end{figure}

\begin{figure*}
    \centering
    \includegraphics[width=0.85\textwidth]{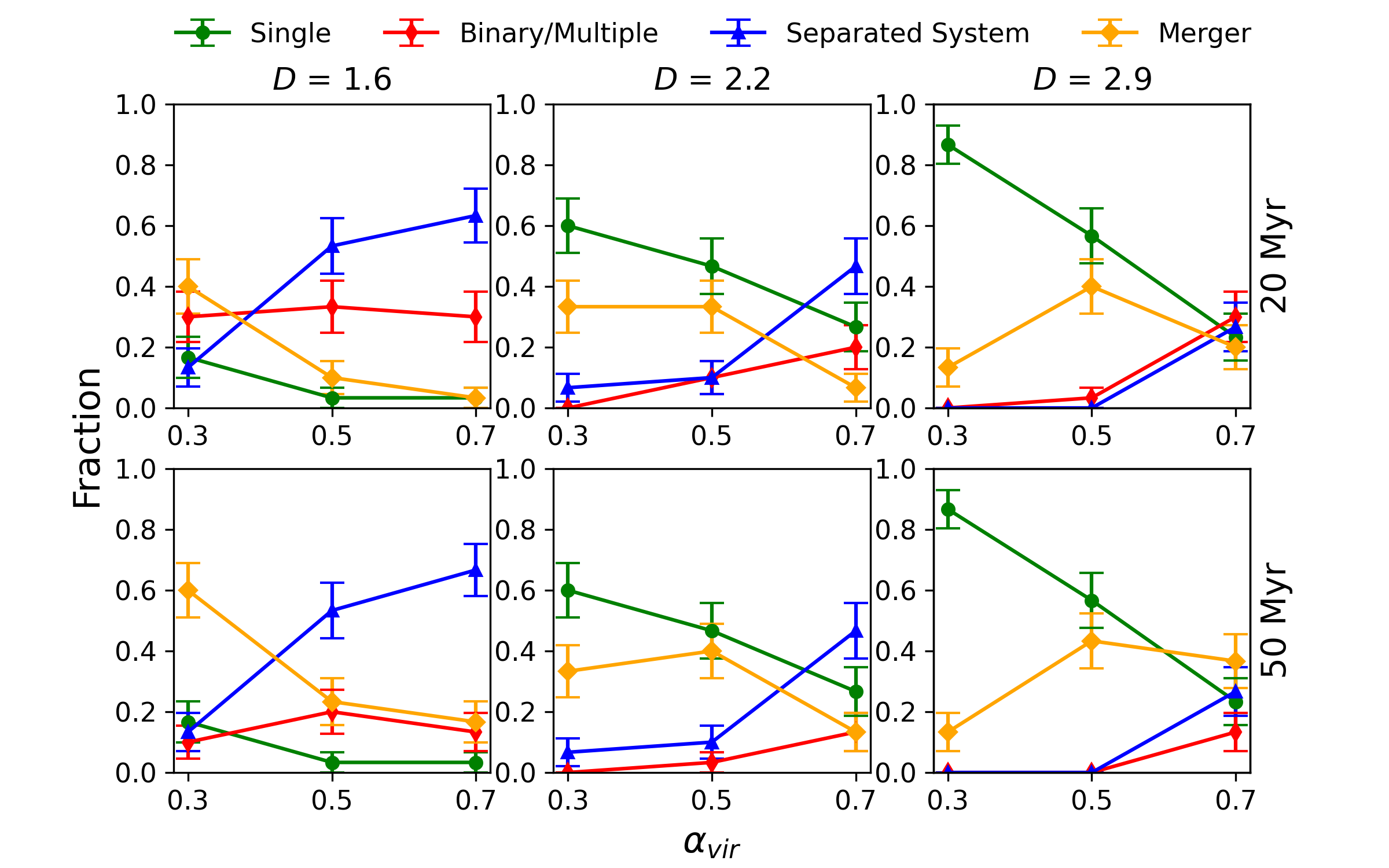}
    \caption{The fractions of the four cluster categories, for the isolated environment. The left, middle, and right panels show the results for the conditions with $D = 1.6$, 2.2, and 2.9, respectively. The top and bottom panels show the fractions at times $t=20$~Myr and $t=50$~Myr, respectively. The different colours represent the fractions for the four different categories.}
    \label{fig:9}
\end{figure*}

We analyse the evolution of the rotation curves of the merger products by comparing the curves at the formation time of the merger product and the end of the simulations at $t=50$~Myr (see Figure \ref{fig:8}). We find that the rotation curve of the merger products do not evolve significantly for the remaining duration of the simulations. The velocity distribution near the cluster center appears to increase up to the cluster's half-mass radius, almost resembles rigid-body rotation. Towards the larger radius, the velocity appears to slightly decrease (top panel of figure~\ref{fig:8}) for several merger products, while the others appear to have relatively flat distributions (middle panel of figure~\ref{fig:8}) and slightly increase distributions (bottom panel of figure~\ref{fig:8}). The initial spatial configuration of the stellar aggregate and its stellar velocity dispersion become the main factors in determining the rotational velocity properties of the merger product. The existence of micro clusters around the merger products is also contributing to the properties of rotation curve.


\subsection{Statistical Analysis in the Isolated Environment}

\begin{figure}
    \centering
    \includegraphics[width=0.45\textwidth]{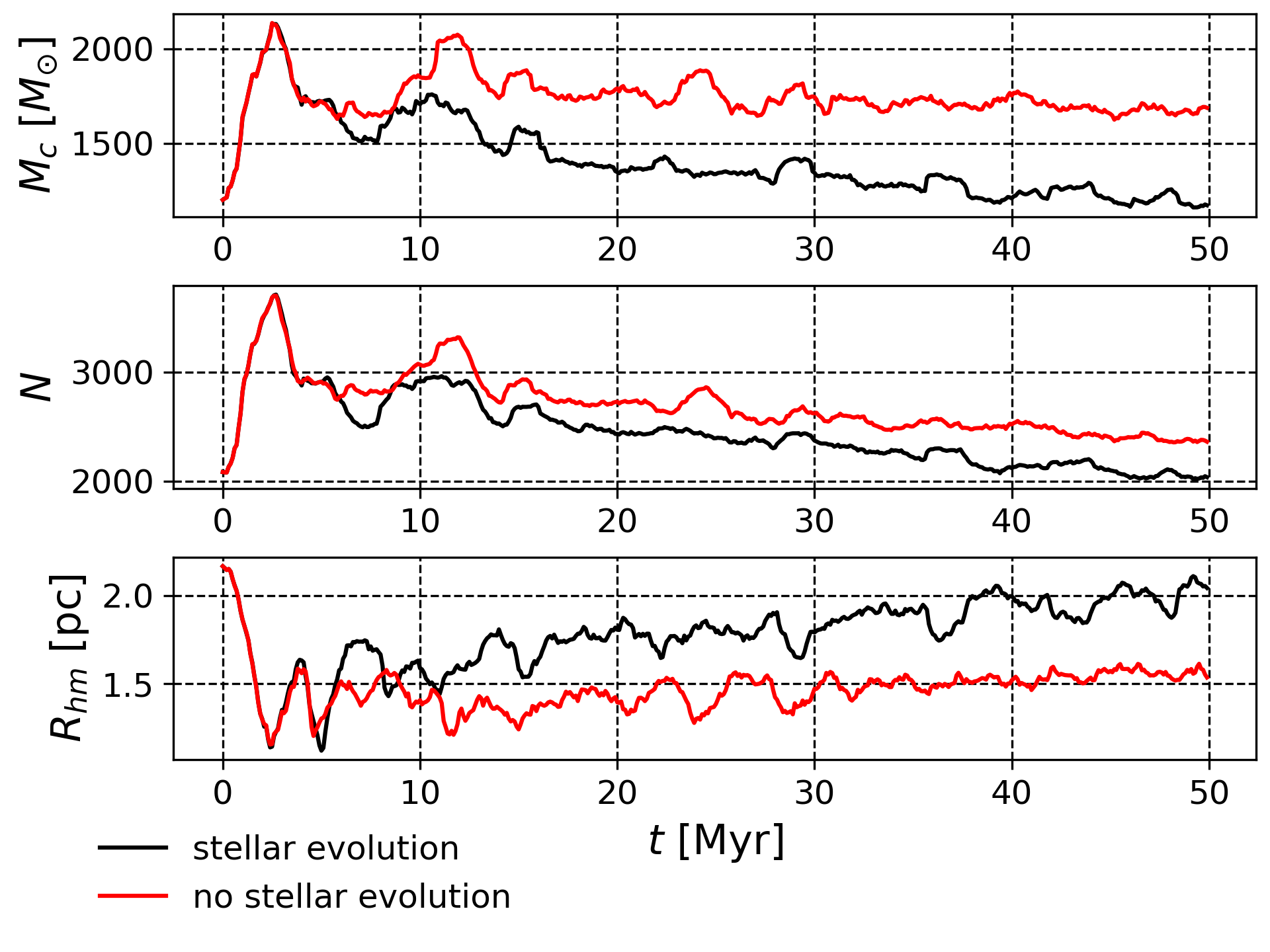}
    \caption{The evolution of cluster's mass, $M_{\rm c}$, number of stars, $N$, and the half-mass radius, $R_{\rm hm}$ of a star cluster, as a function of time, for models with stellar evolution (black curves) and models without stellar evolution factor (red curves), for realisation of a star cluster with initial condition C1 in the isolated environment. The cluster's mass and number of stars are computed within radius of $R = 2~$pc from the cluster's centre, respectively.}
    \label{fig:10}
\end{figure}

\begin{figure}
    \centering
    \includegraphics[width=1.0\columnwidth]{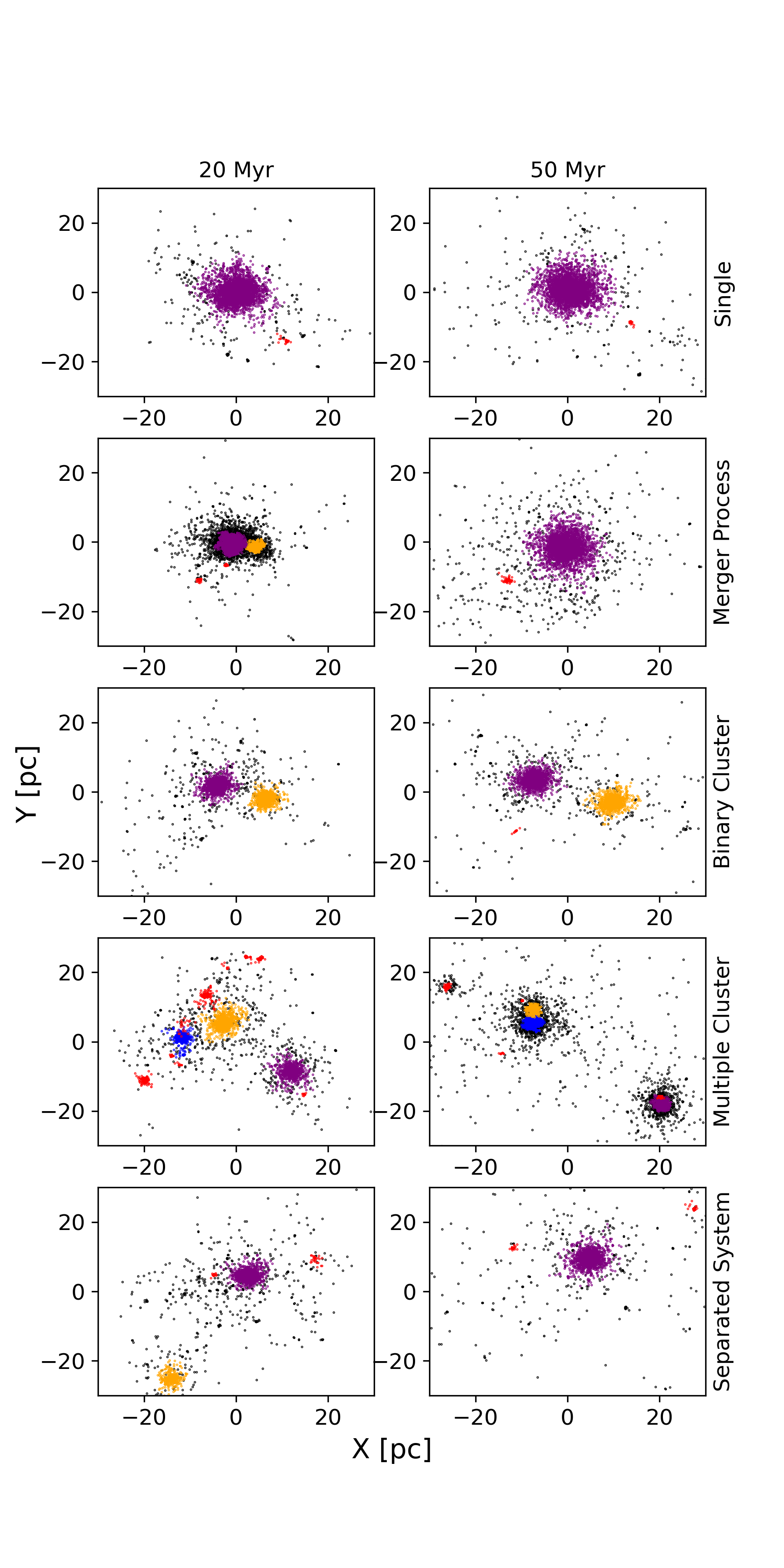}
    \caption{From top to bottom are sequentially samples of single clusters (C5), binary star clusters that have evolved into a merger product (C5), binary clusters (C1), multiple clusters (C4), and separated systems (C2) at times 20~Myr (\textit{left panels}) and 50~Myr (\textit{right panels}). The purple, orange, and blue dots represent the members of the clusters. The members of the micro-clusters are represented with the red dots. As a consequence of the MST algorithm there are also stars which do not clusterize, and these are represented with the black dots.}
    \label{fig:11}
\end{figure}

\begin{figure*}
    \centering
    \includegraphics[width=0.85\textwidth]{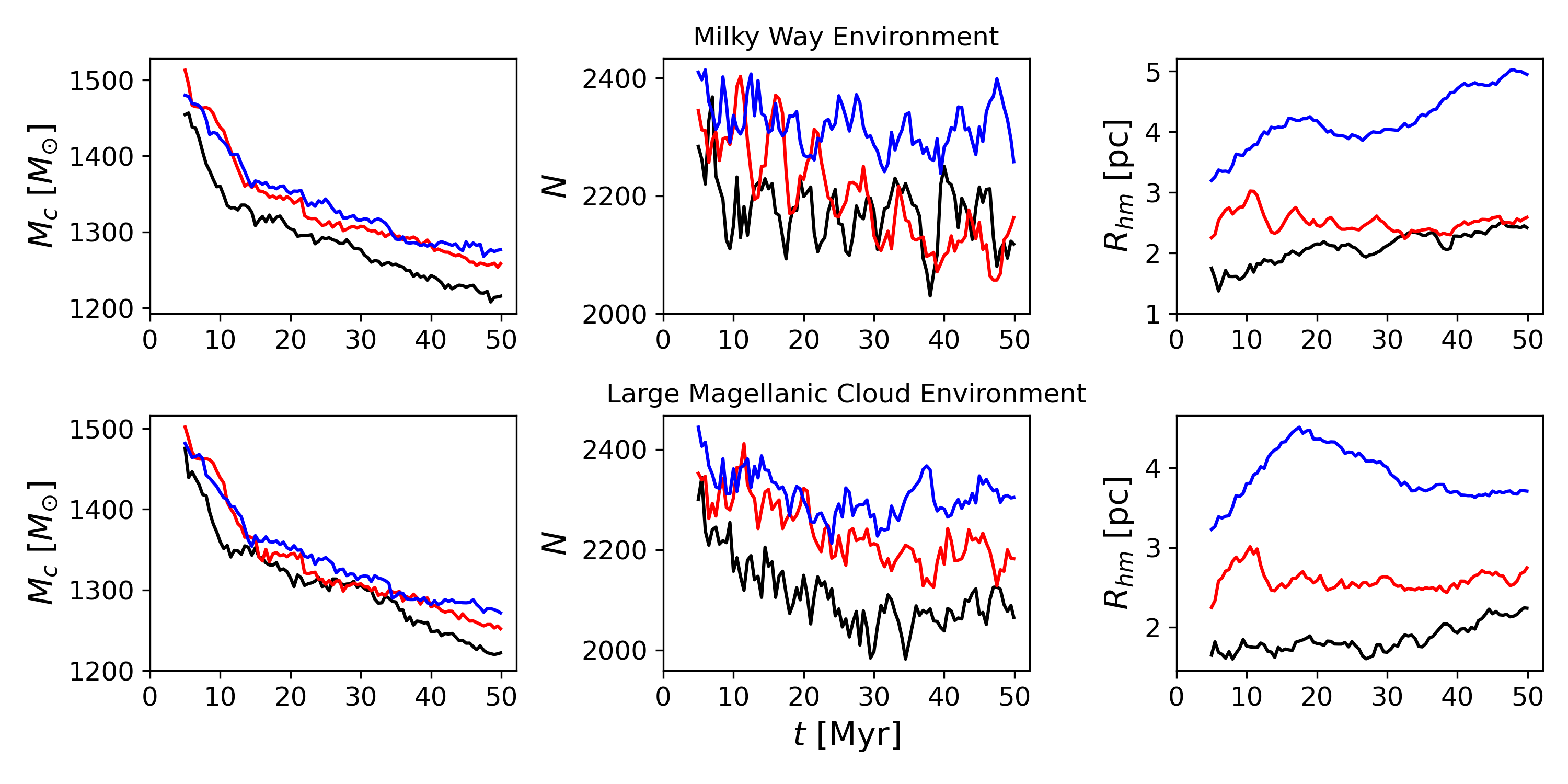}
    \caption{The evolution of the global properties of star clusters in the Milky Way environment (\textit{top panel}s) and LMC environment (\textit{bottom panels}). The black, red, and blue curves show the evolution for representative examples in the ensemble of realisations for models C1, C5, and C9, respectively. The same explanations in Figure~\ref{fig:2} for parameters $M_c$, $N$, and $R_{\rm hm}$.}
    \label{fig:12}
\end{figure*}

\begin{figure*}
    \centering
    \includegraphics[width=0.85\textwidth]{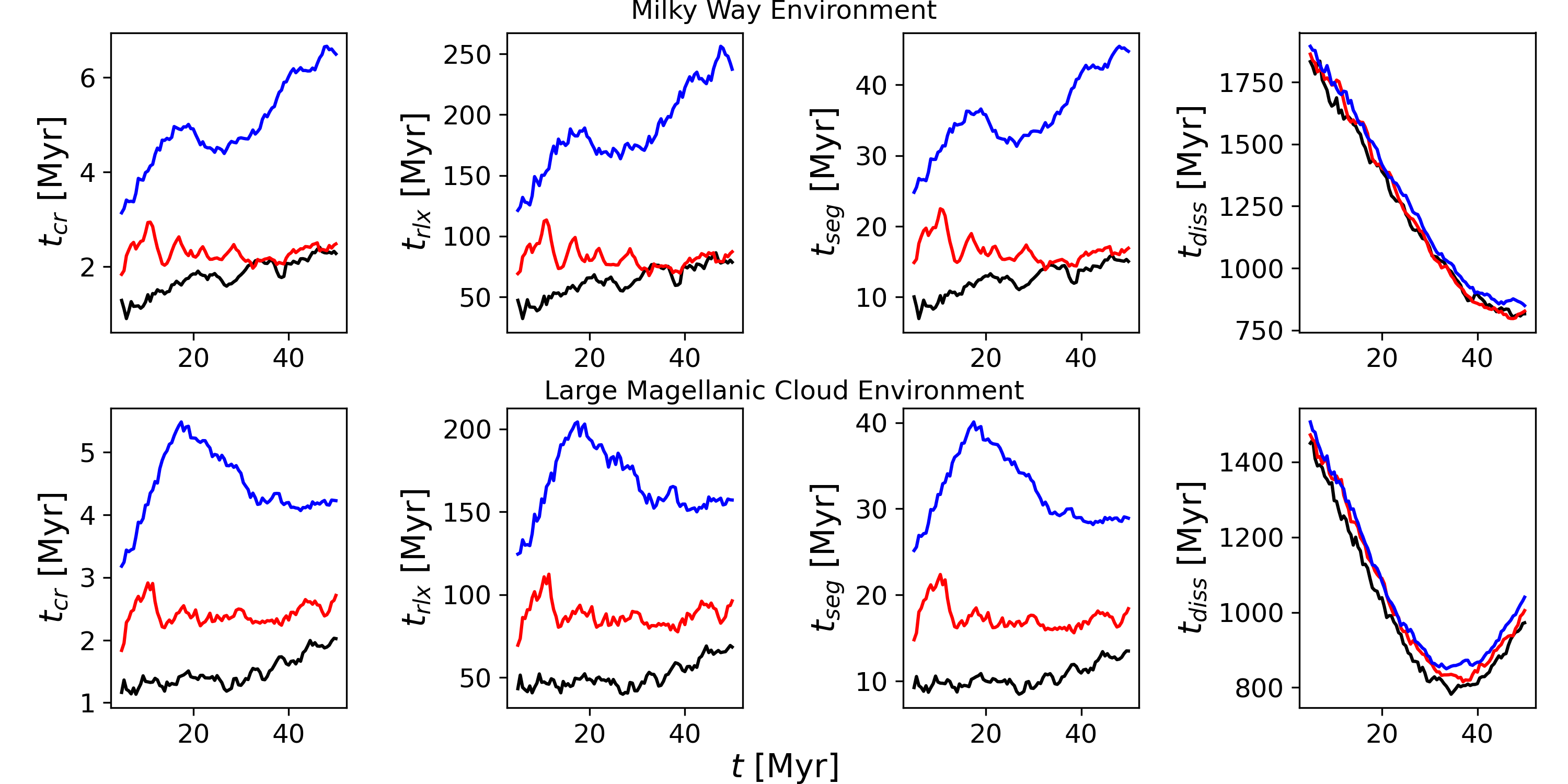}
    \caption{The timescales of star clusters in the Milky Way environment (\textit{top panels}) and in the LMC environment (\textit{bottom panels}). The black, red, and blue curves show the evolution for representative examples in the ensemble of realisations for models C1, C4, and C7, respectively. See Figure \ref{fig:2} for the explanation of parameter $t_{\rm seg}$.}
    \label{fig:13}
\end{figure*}

\begin{figure*}
    \centering
    \includegraphics[width=0.85\textwidth]{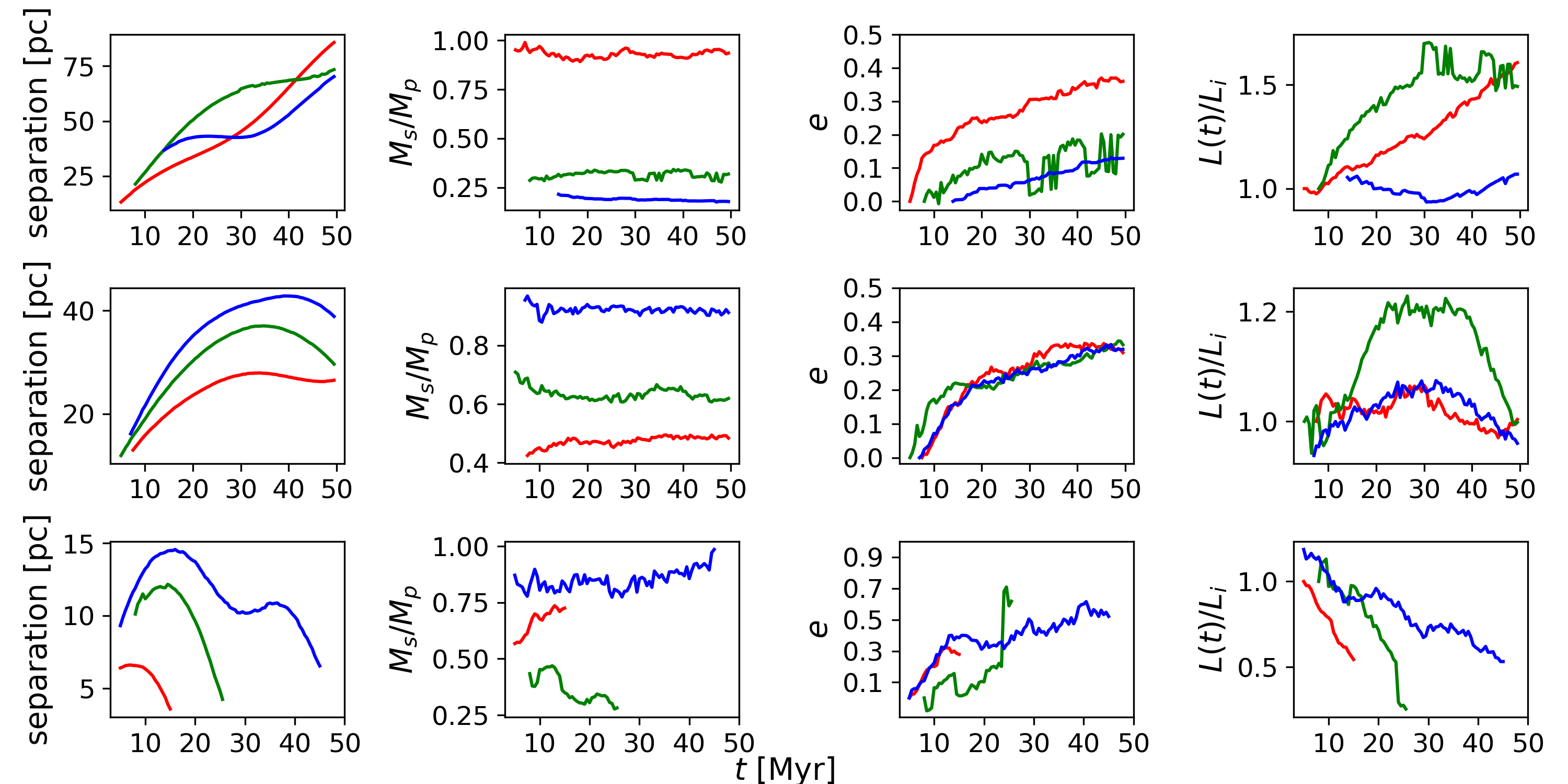}
    \caption{The orbital properties of separated binary cluster (\textit{top panel}), binary cluster (\textit{middle panel}), and merger process (\textit{bottom panel}) in the Milky Way environment. The red, green, and blue curves in the top panels represent the dynamical evolution for initial conditions C5, C7, and C8, respectively. In the middle panels, the curves  represent initial conditions C1, C5, and C8, respectively, while the colours in the bottom panels represent the results for the models with initial conditions C1, C4, and C8, respectively.}
    \label{fig:14}
\end{figure*}

\begin{figure}
    \centering
    \includegraphics[width=0.45\textwidth]{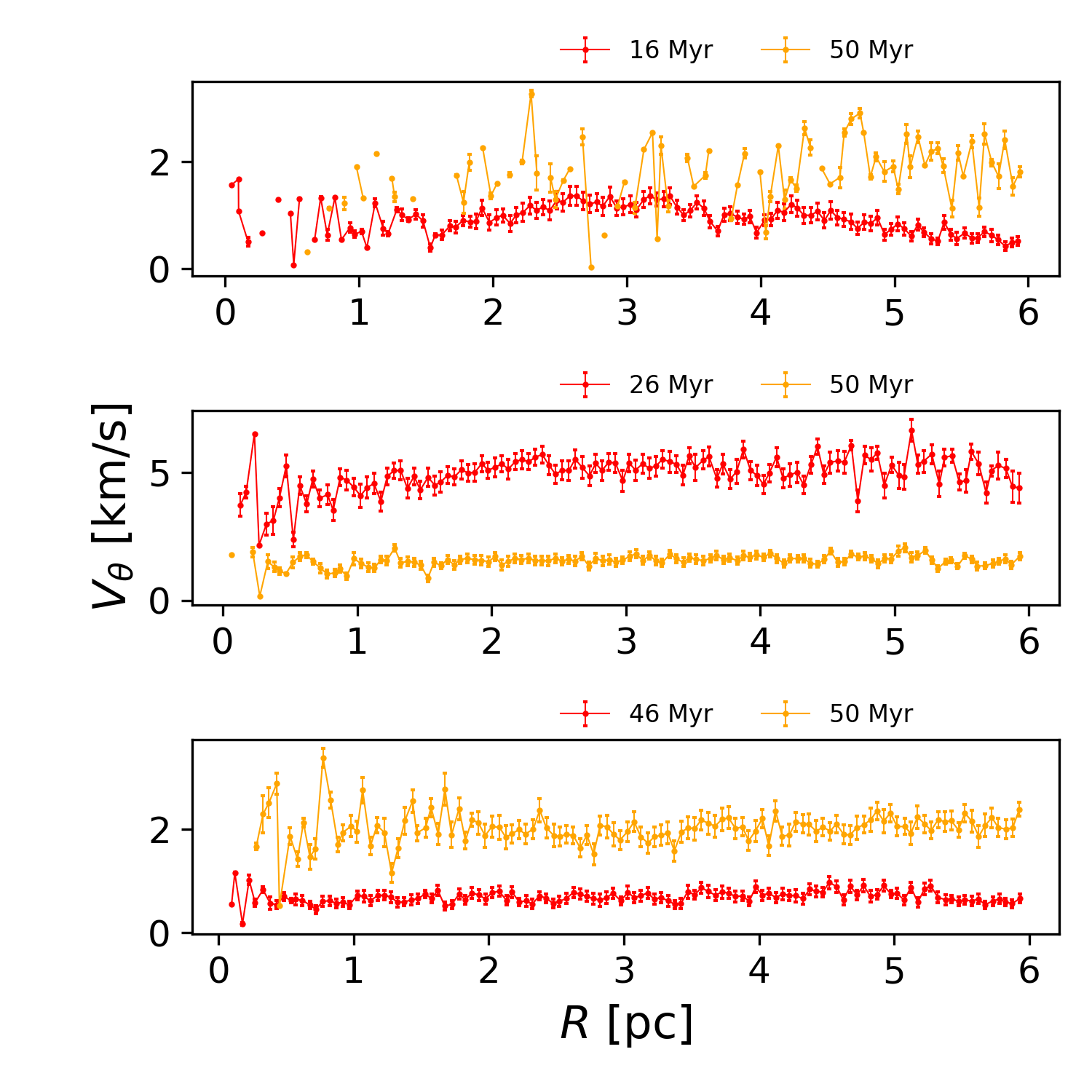}
    \caption{The rotation curves of merger products for the Milky Way environment with initial conditions C1 (\em top), C4 (\em middle), and C8 (\em bottom).}
    \label{fig:15}
\end{figure}

In Figure~\ref{fig:9} we present the distribution of the clusters over the four different dynamical categories, for the isolated environment. 
Stellar aggregates with a higher degree of initial substructure (i.e., smaller $D$) have a higher probability of evolving into binary/multiple clusters, due to tendency of the clumps of stars to gravitationally bind shortly after formation.

A higher initial virial ratio ($\avir$) results in a higher probability of forming a separated system. Such systems tend to have a somewhat higher net angular momentum shortly after formation. Consequently, clustered groups of stars (as well as individual stars) have a general tendency to move away from each other. It is still possible for high-$\avir$ systems to form binary/multiple clusters, although it is more likely for the system to evolve into a separated system.

The fraction of binary/multiple systems decreases as the clusters grow older, due to the gravitational interactions between the clusters, which can trigger a merging event. This explains why the fractions of merged clusters increases over time. A combination of mass loss and angular momentum conservation also triggers several binary/multiple clusters to evolve into separated systems. This process contributes to a growing fraction of separated systems as time passes.

\begin{figure*}
    \centering
    \includegraphics[width=0.85\textwidth]{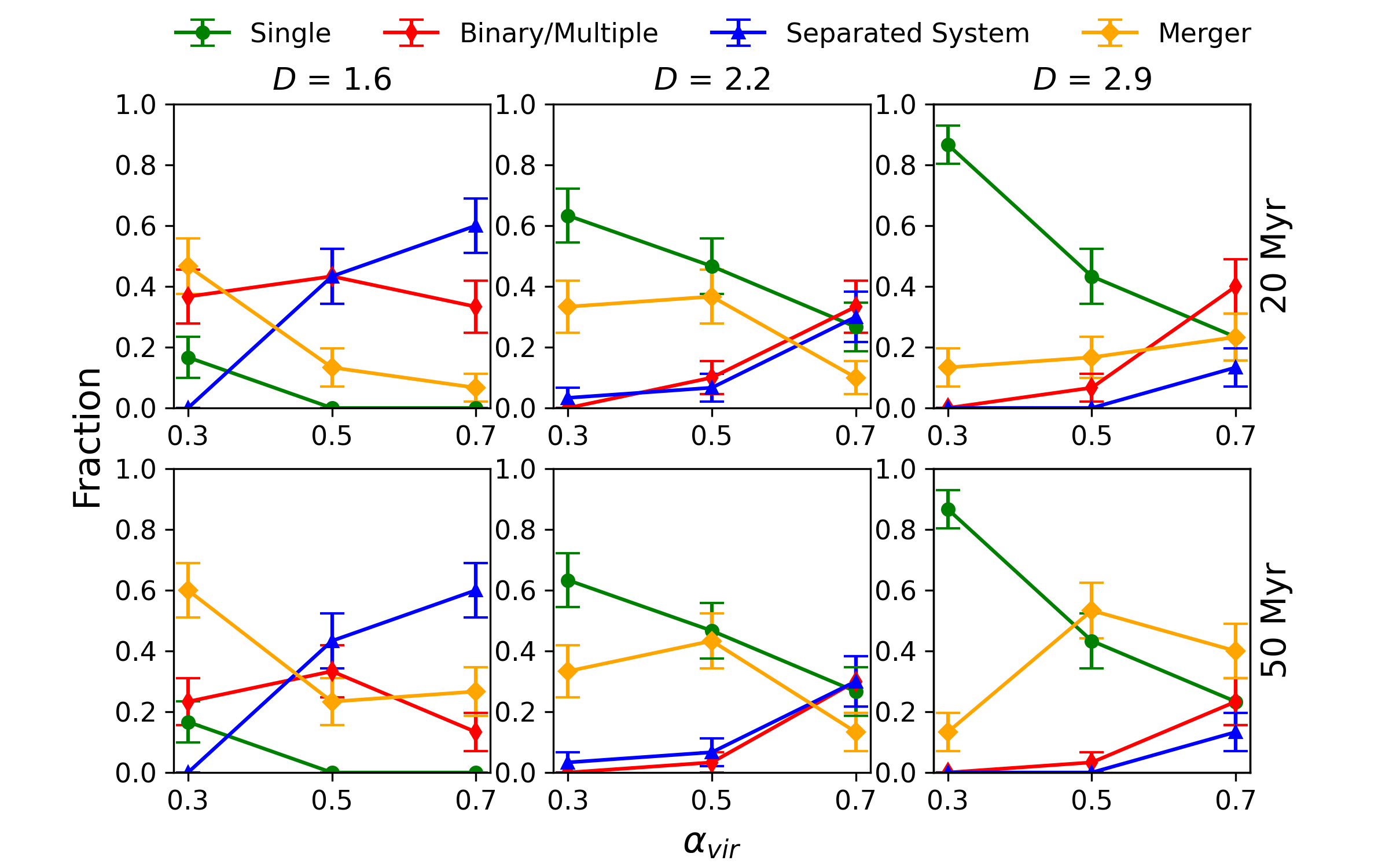}
    \caption{The fraction of single clusters, binary/multiple clusters, separated systems, and merger products in the Milky Way environment. Panel descriptions are as in Figure~\ref{fig:9}.}
    \label{fig:16}
\end{figure*}

\begin{figure*}
    \centering
    \includegraphics[width=0.85\textwidth]{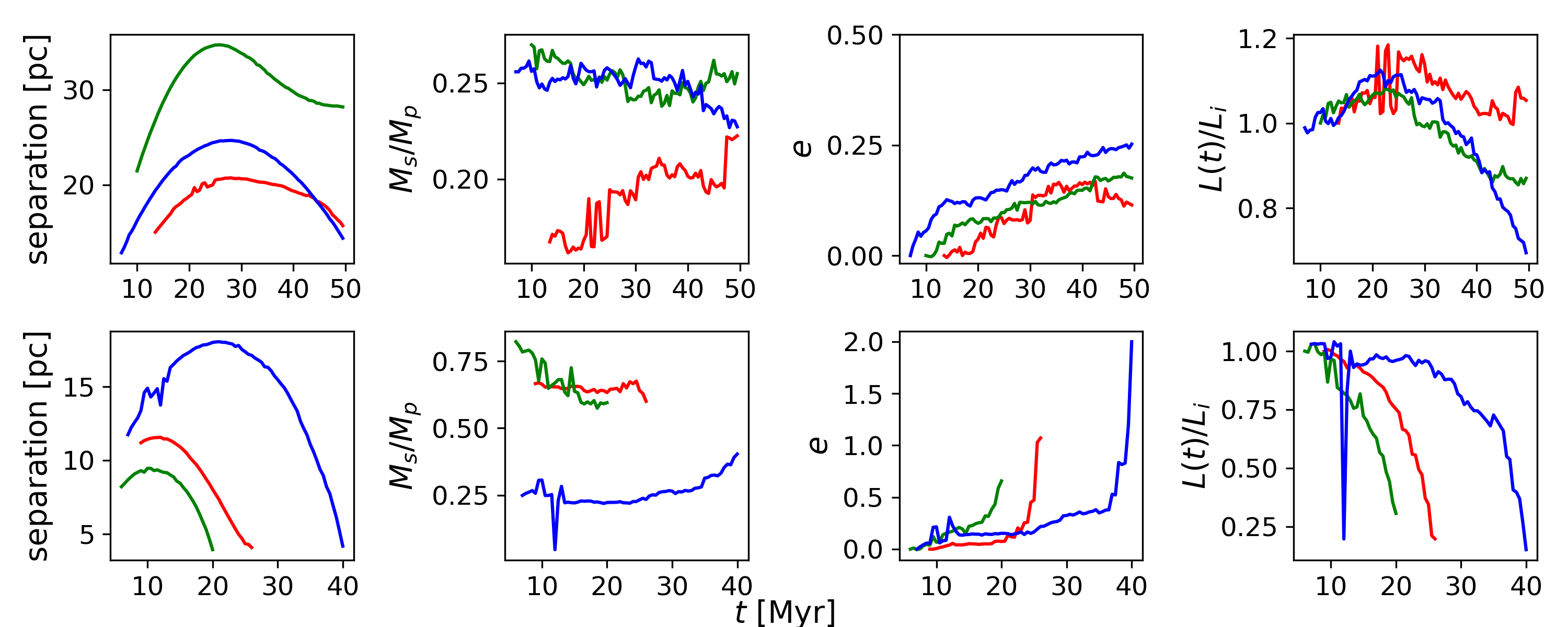}
    \caption{The orbital properties of binary star cluster (\textit{top panels}) and merging clusters (\textit{bottom panels}) in the LMC environment. The red, green, and blue curves in the top panel represent the dynamical evolution for initial conditions C1, C5, and C9, respectively. In the bottom panels, the red, green and blue curves represent initial conditions C1, C5, and C8, respectively.}
    \label{fig:17}
\end{figure*}

\begin{figure}
    \centering
    \includegraphics[width=0.45\textwidth]{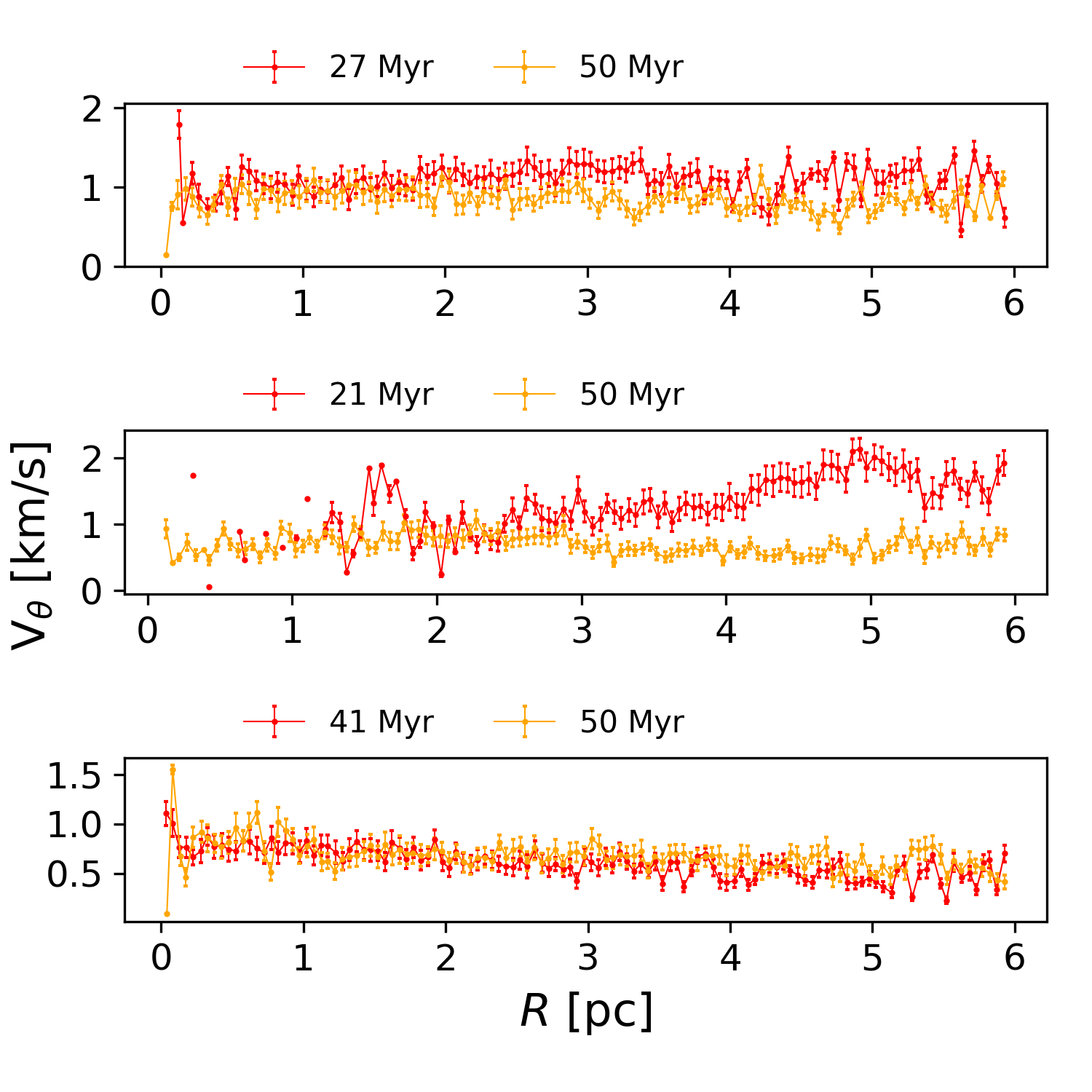}
    \caption{The rotation curves of merger products for the LMC environment with initial conditions C1 ({\em top}), C4 ({\em middle}), and C8 ({\em bottom}).}
    \label{fig:18}
\end{figure}

\begin{figure*}
    \centering
    \includegraphics[width=0.85\textwidth]{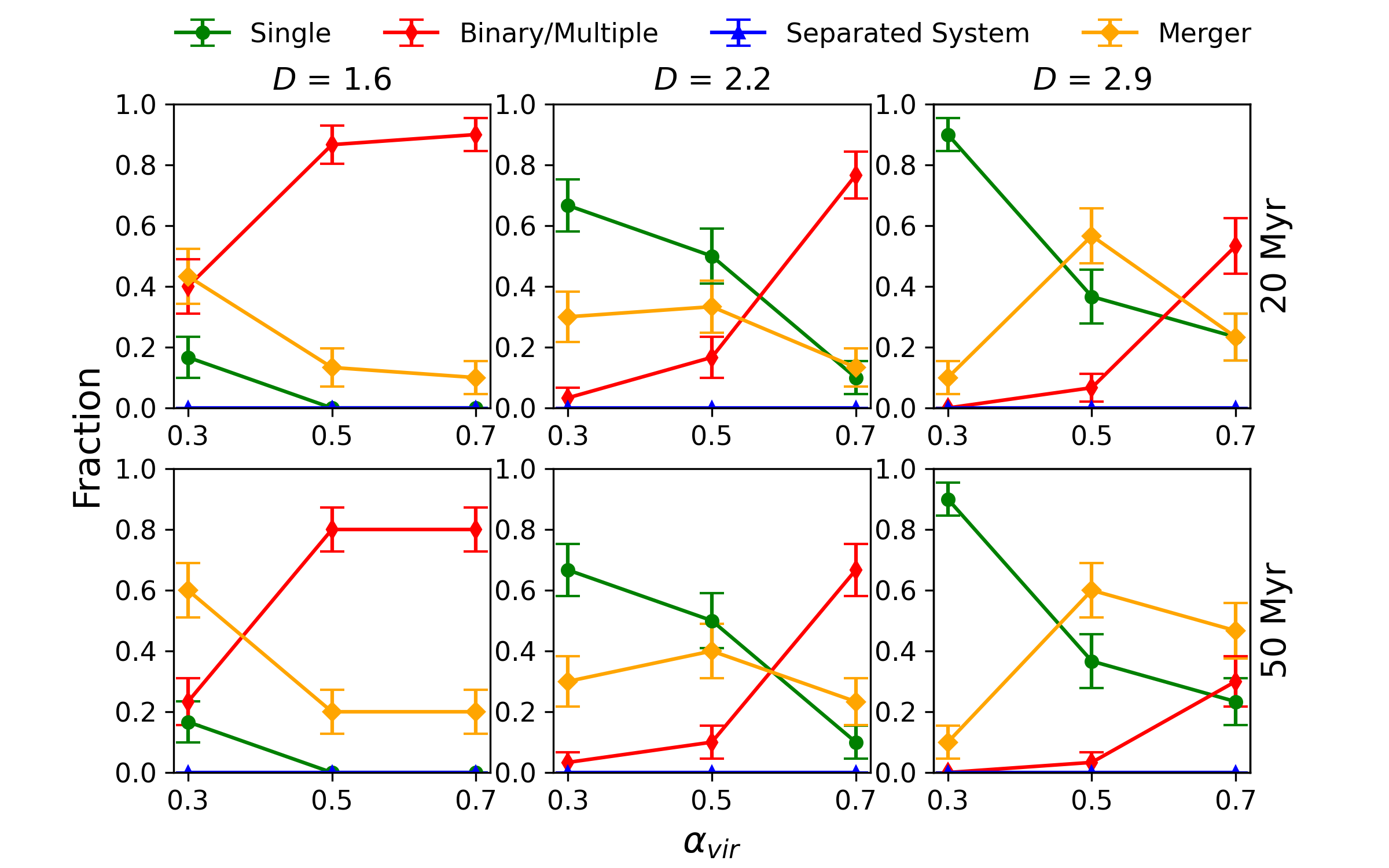}
    \caption{The fraction of single clusters, binary/multiple clusters, separated systems, and merger products in the LMC environment. The panel descriptions as explained in Figure \ref{fig:9}.}
    \label{fig:19}
\end{figure*}

The most likely initial conditions that trigger the formation of binary/multiple clusters occur for models with small initial values for $D$ (for any $\avir$). Here, the fraction of binary/multiple clusters is $\sim$30\% at time $t=20$~Myr, which decreases to $\sim$15\% at time $t=50$~Myr. The fractions obtained at time $t=20$~Myr are somewhat different from those obtained by \cite{Arnold2017} due to the different initial masses and initial numbers of stars used in their models. Furthermore, stellar evolution is included in our simulations, which tends to contribute to the formation rate of binary star clusters. In general, the star clusters in our simulations evolve somewhat faster than those studied in \cite{Arnold2017}.

We do not expect a significant change in the fraction of single clusters between times $t=20$~Myr and $t=50$~Myr, since single clusters are typically formed at early times. The highest fraction of single clusters occurs in stellar aggregates that initially have a more or less homogeneous structure ($D=2.9$), for all choices of $\avir$. Mergers between the two members of a binary star cluster do occur, though. From an observational perspective, these are single star clusters, although we categorise these separately as merger products in our analysis. These merger products form at all times, increasing the fractions of mergers from $\sim 40\%$ at $t=20$~Myr to $\sim 60\%$ at $t=50$~Myr. The other evolutionary fate of binary star clusters is a separation into two gravitationally unbound star clusters. The fraction of separated systems remains mostly constant over time, indicating formation at early times. In the absence of an external tidal field, the fraction of binary star clusters evolving into separated systems is much much smaller than the fraction of stellar aggregates evolving into single clusters through a merger process.

In order to understand the contribution of stellar evolution to the evolution of the systems, we compare otherwise identical models with and without stellar evolution in Figure~\ref{fig:10}. Stellar evolution enhances mass loss, and also results in a higher escape rate of stars from the system, as the gravitational potential decreases as the clusters age. The increased mass loss from the stellar evolution also speeds up the process of cluster's expansion. These trends are present in all the models. Figure~\ref{fig:11} shows a representative sample of systems in each of the categories, at times $t=20$~Myr and $t=50$~Myr. Most  micro-clusters are present at $t\approx 20$~Myr. These micro-clusters usually merge with their parent clusters before $t=50$~Myr. Several micro-clusters escape from the system, particularly for systems with high values of $\avir$.


\subsection{Comparison with the Milky Way Environment}

Perturbations by an external galactic tidal field strongly affect the dynamical stability of binary star clusters. In the top panels of Figure~\ref{fig:12}, we present the impact of an external tidal field on the dynamical evolution of the evolution of the mass, the number of stars, and the size of the clusters. In a Milky Way environment, the modelled clusters lose $\sim 15-20\%$ of their member stars within 50~Myr. Within the same time, their total mass decreases by $\sim 20-30\%$, due to stellar escapers and stellar evolution. This results in the star clusters to gradually dissolve, and the instantaneous value of $t_{\rm diss}$ decreases over time (see the top panels of Figure~\ref{fig:13}). These findings are similar to those of the isolated environment, and  indicate that the presence of a tidal field does not significantly affect the internal dynamics of star clusters within this time span.

Figure~\ref{fig:14} shows the orbital properties of binary star clusters and separated systems in the Milky Way environment. The separations between the two star clusters for separated systems (top panels) are somewhat smaller than similar models in the isolated environment. Most of separated systems have mutual separations less than 60~pc at $t = 20$~Myr and separations that do not exceed 100~pc at $t = 50$~Myr. Just like in the isolated environment, the resulting mass ratios of the binary star clusters vary substantially, with only small fraction of the binary star clusters obtaining a mass ratio near unity. The latter occurs most frequently for substructures with $D=2.2$. The median values of the mass ratios for initial substructures of $D=1.6$, $D=2.2$, and $D=2.9$ at $t=20~$Myr are $q=0.36$, $q=0.75$, and $q=0.36$, respectively. As in the isolated environment, the mass ratios have also slightly evolved at time $t = 50~$Myr. Moreover, our study shows that binary star clusters can reach maximum separations of $\sim 50$~pc before they start to approach each other (Figure~\ref{fig:14}; middle panels).

Binary star clusters that merge into a single cluster tend to have smaller maximum separations: $\leq 20$~pc before merging. This upper limit is somewhat larger than in the isolated environment, in which the maximum separations do not exceed 13~pc. The merger products in the Milky Way environment have different kinematic properties when compared to those formed in the isolated environment. After the formation of the merger products, the degree of global rotation increases over time for some, while it decreases for others (see Figure~\ref{fig:15}). It indicates that the star clusters will reach their stability in longer lifetime than 50~Myr. 
Furthermore, we find the rotation velocity curves tend to flatten at larger radii as time passes, as a consequence of internal dynamical evolution and interaction with its environment, as suggested by \cite{Kim2002} and \cite{Priyatikanto2016}. However, several merger products have initial rotation curves that drop off at larger radii (see, e.g. \citealt{Kim2004}; \citealt{Kim2008}; \citealt{Hong2013}). The rotation speed then increases again at the outer radii as the cluster evolves. This indicates that over time, the dynamical stability of star clusters is affected by the presence of the galactic tidal field. According to \cite{Priyatikanto2016}, the dynamical processes lasting in the binary clusters before undergoing a merger, could be the strong factors of this difference.

The distribution of the systems over the four categories in Figure \ref{fig:16} shows that the fraction of binary/multiple in the Milky Way environment can reach $\sim 45\%$ at  $t=20$~Myr and $\sim 30\%$ at $t=50$~Myr, for any $\avir$. The fraction of separated system at $t=20$~Myr is approximately 60\%, and no new separated systems form after that. The fraction of binary/multiple systems is higher than that in the isolated environment, while the fraction of separated system is similar. The fraction of merger products in the Milky Way environment is $\sim 50\%$ at $t=20$~Myr and increases to be $\sim 60\%$ at $t=50$~Myr.

In summary, we find several different pathways for the dynamical evolution between star clusters, depending on whether they evolve in isolation, or whether they are embedded in the Milky Way environment:
\begin{enumerate}
    \item Several of the systems (with identical initial conditions) become separated systems in the isolated environment, while they evolve into binary/multiple clusters in the Milky Way environment.
    \item Several of the systems that become single clusters  in the isolated environment evolve into binary star clusters in the Milky Way environment, and merge shortly after.
    \item A few of the binary/multiple clusters that appear in the isolated environment, evolve into separated systems when placed in the Milky Way environment.
\end{enumerate}
These different pathways of dynamical evolution can occur due to the direct or indirect effect of the Galactic tidal field. In addition, these pathways depend also strongly on the orientation of the system with respect to the Galactic centre \citep{Priyatikanto2016}.


\subsection{Comparison with the LMC Environment}

During their first 50~Myr, star clusters in the LMC environment lose $\sim 10\%$ of member stars, and they lose $\sim 10\%-25\%$ of their initial total mass. These changes are comparable to our findings for the Milky Way environment and for clusters that evolve in isolation. Variations in the dynamical timescales at which the different processes occur are a consequence of the physical properties of the clusters that form from the substructured initial conditions (bottom panel in Figure~\ref{fig:13}). The tidal field of the LMC has perturbed the stability of star clusters and causes the clusters to undergo gradual dissolution during the 50~Myr with smaller value of $t_{\rm diss}$ as the cluster grows older. The increasing dissolution timescale at time $t \ga 35$~Myr depends on the position of star cluster in which it moves away from the galactic centre, so that the effect of galactic tidal field to the star cluster becomes smaller and the dissolution rate becomes slower.

We do not find any separated systems in the LMC environment, both at $t=20$~Myr and at $t=50$~Myr. All binary/multiple clusters are gravitationally-bound, with mutual separations less than of $\sim 35$~pc (see the top panels in Figure~\ref{fig:17}). This upper limit for the separation is smaller than in the Milky Way environment, but comparable to the maximum separations seen in the isolated environment. The mass ratios of binary clusters in the LMC environment do not differ significantly from those of the isolated environment at $t=20~$Myr and $t=50~$Myr. The median values of the mass ratios for initial substructures with $D=1.6$, $D=2.2$, and $D=2.9$ at $t=20~$Myr are $q=0.36$, $q=0.78$, and $q=0.24$, respectively. Merger products are formed at all times in the LMC environment, after the collision of the two members of a binary star cluster which have maximum separations smaller than $\sim 20$~pc (see the bottom panels in Figure \ref{fig:17}). In addition, we find that merger products are commonly formed after a close encounter between two companions, when they approach each other nearly head-on and in a retrograde binary cluster orbit.

The rotation curves of the merger products in the LMC environment are comparable to those in the Milky Way environment, where the curves are more or less flat at large radii (see Figure~\ref{fig:18}), while several merger products show increased velocities at larger radii. At $t=50$~Myr, several merger products with radially decreasing rotation curves remain, such as those in the isolated environment. We find that the rotation velocities in the LMC environment are similar to those of the merger products in the isolated environment, but somewhat smaller than those of the merger products in the Milky Way. We attempt to investigate the reason behind this result. Rotation was found to accelerate the star cluster's dynamical evolution by transferring the angular momentum outwards, by speeding up the process of mass segregation, and by reducing the core collapse timescale (\citealt{Kim2002}, \citealt{Kim2004}, \citealt{Kim2008}). The presence of a galactic tidal field in our study will also speed up the transfer of angular momentum outward, as stars in the outer regions of the cluster are stripped off. This accelerates the internal dynamics of the cluster (two-body relaxation, mass segregation, core collapse, etc; see, \citealt{Mapelli2017}). This results in increased rotation speeds with time. As explained in Section~\ref{section:tidalmodels}, the tidal radius of a cluster is roughly 43.5\% larger in the Milky Way environment than in the LMC environment. The tidal field of the LMC  results in rotation curves should be expected more pronounced than those in the Milky Way environment. However, there seems to be another factors affecting the lower rotation curves significantly in the LMC. We then also calculate the angle between each cluster's rotation axis and its orbital plane in the galaxy. Most of the LMC merger products that have lower rotation curves than those in the Milky Way environment are found to have a nearly straight angle. These clusters appear to be more robust against perturbations from the galactic tidal field. Thus, a possible reason of the cluster’s lower rotation speed are more likely the angle between cluster's rotation axis and its orbital plane in the galaxy.

The typical pathways for the evolution of the stellar aggregates are similar to those evolved in the Milky Way environment. When compared to the Milky Way environment and the isolated environment, the fraction of binary/multiple cluster in the LMC environment are highest (see Figure~\ref{fig:19}). The fraction reaches a maximum value of $\sim 90\%$ at $t=20$~Myr and decrease to $\sim 80\%$ at $t=50$~Myr. This suggests that the systems in the LMC environment can survive up to 100~Myr or even longer, as suggested by \cite{Priyatikanto2019b}. The fraction of merger products in the LMC higher:  $\sim 3\%-10\%$ at $t=50$~Myr. In the LMC environment, we find that most binary star clusters ultimately merge, rather than forming a separated system. The fraction of merger products in the LMC environment is somewhat higher than in the Milky Way environment. This is particularly the cases for models with $D = 2.9$ and $\avir = 0.5-0.7$ (see Figure \ref{fig:19}), which mostly evolve into separated systems in the Milky Way environment, while in the LMC environment these mostly end up in  mergers. The simulations for such models show that as two binary cluster companions are receding and the system moves near to the galactic plane, the separation of the companions decreases rapidly, which results in a merger product, instead of remain in a separated system, as in the Milky Way environment. However, in \cite{Priyatikanto2016}, binary star clusters near the Milky Way plane are less likely to merge; rather they tend to evolve into separated systems, as a consequence of the limited range of a cluster's separation to produce merger process (see, e.g., figure~2 in \citealt{Priyatikanto2016}). The results obtained by \cite{Priyatikanto2016} are affected by the  Milky Way's tidal field. The different tidal field and structures in the LMC might contribute to provide different possible range of cluster's separation to produce merger process with small inclinations and various values of ascending node in the LMC environment.


\section{Discussion and Conclusions}
\label{sec:4}

We have performed gravitational $N$-body simulations to investigate the formation process and subsequent dynamical evolution of binary star clusters. We have investigated the process of how stellar aggregates evolve with different initial conditions (substructure parameter $D$ and virial ratio $\avir$), and in different environments: (i) in isolation; (ii) in the Milky Way environment, and (iii) in the LMC. Star clusters with a high degree of initial substructure (i.e., small $D$) are more likely to form binary star clusters. Star clusters with a higher stellar velocity dispersion (higher $\avir$), on the other hand, are more likely to form a separated systems (i.e., gravitationally unbound star clusters). 

At early times, the collapse of a stellar aggregate will result in a modest degree of mass segregation ($\Lambda \approx 1-3$). Binary star clusters then tend to form at $t \ga 5$~Myr, after a short phase of violent relaxation. In some cases, multiple star clusters are formed, notably in the simulations which are initially supervirial ($\avir = 0.7$). Such multiple systems are generally short-lived due to the high rate stellar escapers and escaping mini-clusters, and rapidly evolve into binary star clusters. The gravitational interactions between the components results in binary star cluster to merge into larger rotating star clusters when the separation between the two clusters is sufficiently small. Binary star clusters with a larger separation, on the other hand, tend to widen their orbits and subsequently separate into two star clusters that move away in opposite directions.

We find that the fraction of binary/multiple clusters in the Milky Way environment is typically $\sim 45\%$ at $t=20$~Myr, and $\sim 30\%$ at $t=50$~Myr, with separations less than $\sim 50$~pc. On the other hand, most of the binary star clusters with orbital separations less than $\sim 20$~pc tend to merge. Gravitational interactions between star clusters can lead to a merger process through the spiral-in of orbits with smaller eccentricity ($e<1$) and through highly eccentric, head-on collisions ($e \geq 1$). A fraction $\sim 50\%-60\%$ of merger products is found in the Milky Way environment during the first 50~Myr. Approximately $60\%$ of the binary star clusters with separations above $\sim 60$~pc evolve into separated systems at later times.

In the LMC, $\sim 80\%-90\%$ of the stellar aggregates evolve into binary star clusters at some stage during the 50~Myr, and form with separations less than $\sim 35$~pc. About $\sim 10\%-60\%$ of the merger products are formed from binary star clusters, all of which have with separations less than $\sim 20$~pc. No separated systems are formed in the LMC environment during the 50~Myr that they are evolve, for any of the sets of initial conditions.

The dynamical stability and evolution of star clusters is also affected by a galactic tidal field, and therefore its dynamical fates (single cluster, binary star cluster, separated system, or merged system) is determined by the tidal field. The tidal field also affects the rotation curves of merger products at large radii. We have measured the (evolving) angle between the rotation axis of each cluster, and the  orbital plane of the cluster in the galaxy. We find that merger products with evolving rotation curves exhibit changes in their orientation angles. As the angle decreases, the merger product's rotation curve increases more significantly. We also find several merger products in the Milky Way and LMC environments which have relatively stable rotation curves, which occurs especially when the angle is perpendicular. The latter is also commonly identified in the isolated environment. 
\cite{Kacharov2014} and \cite{Priyatikanto2016} have shown that the rotation axis of a cluster with respect to its orbital plane can strongly affect the evolution of the rotation curves. \cite{Kacharov2014} showed there is an oscillation of cluster's rotation velocity as the rotation axis changes as a cluster orbits around the galactic centre. However, our simulations are limited to a total integration time of 50~Myr, which is insufficient to observe such long-term effects.

Among the merger products in the Milky Way environment, typically $\sim 37\%$ change rotation curves significantly within the simulated timespan of 50~Myr. The corresponding fraction is somewhat smaller in the LMC environment: $\sim 34\%$. The changes in rotation velocities of the merger products are mostly above 1~km\,s$^{-1}$, between the formation time of the merger product and the total simulation time of 50~Myr. However, some merger products have small changes of rotation velocities i.e., less than 0.5~km\,s$^{-1}$; these smaller variations are mostly within the spread of the cluster's rotation velocities. The star clusters typically lose $\sim 20\%-25\%$ of their initial total mass during 50~Myr, and $\sim 10\%-15\%$ of the member stars from each system. The external tidal field does not appear to strongly affects the evolution of the total mass of the bound system, the number of member stars, or the radii of the star clusters that form. Stellar evolution and the dissipation of angular momentum, on the other hand, are more important factors during this timespan.

The probability of formation of binary star clusters, as well as their dynamical fate, is primarily affected by the external tidal field in different ways. The tidal radius and strength of the galactic tidal field experienced by the system depend on its position with respect to the galactic centre. Mutual disruption will be dominant and cannot be neglected when the separation of two star clusters is less than three times their tidal radii \citep[see, e.g.,][]{Innanen, BinneyTremaine, Marcos2009}. In Section~\ref{section:tidalmodels}, we have shown that the initial tidal radius of star clusters in the LMC is smaller than that of clusters in the Milky Way. This indicates that the frequency of mutual disruptions in the LMC star clusters is larger than in the Milky Way. Moreover, for the positions in the galaxies at which the clusters in this study were initialised, star clusters in the LMC feel a stronger tidal force than those in the Milky Way. Under these conditions, the LMC's tidal field less disruptive when compared to the contribution of mutual disruption of the two star clusters in the system. In the Milky Way environment, on the other hand, the formation process of binary star clusters are affected by the galactic tidal field. Several separated systems are pulled back by the tidal field into a bound state, or alternatively, retained several systems in a separated systems with higher mutual separations.
Such processes can occur due to the initial spatial orientation of binary star clusters with respect to the galactic centre, as suggested in \cite{Priyatikanto2016}. Many of our star clusters appear to have initial orientations that are beneficial for the formation of binary star clusters in the LMC. This process should also have occurred in the Milky Way, but its stronger tidal field has triggered separations and mergers.

Our results are somewhat different from those of \cite{Arnold2017}, due to (i) the higher initial mass of the clusters modelled in our study, and (ii) our inclusion stellar evolution. Both these differences result in a faster dynamical evolution of the clusters in our study. On the other hand, we have not considered the effect of presence of binary stars in our study. Previous studies have shown that many binary stars are formed in the Galaxy, and these binary stars affect the internal evolution of star cluster, for example through preventing core collapse, and through slowing down the dissolution process of star cluster (e.g., \citealt{deGrijs}; \citealt{Kouwenhoven}; \citealt{Martinez-Barbosa2016}). Moreover, the \mistix{} survey showed that many of the youngest star clusters are surrounded by gas. This gas  experiences ram pressure, which decreases the rate of dynamical evolution. Further studies should include the effects of both primordial binary stars and the presence of gas. More interestingly is to investigate the influences of the star cluster's initial mass and initial number of stars on the probability of forming binary star clusters. The effect of a non-axisymmetric external gravitational potential, such as spiral arms and bar structures, should also contribute in the dynamical evolution of star clusters \citep[e.g.,][]{Mishurov2011, Martinez-Barbosa2016, Martinez-Barbosa2017}. Besides that, the availability of the early third data release of GAIA will provide a new challenge of cluster membership determination and of cluster physical properties estimation. These data can be compared with results from $N$-body simulations to determine the physical status of candidate cluster pairs, and to investigate cluster pair formation in the Milky Way and in the LMC (e.g. \cite{Kovaleva2020}).


\section*{Acknowledgements}
We are grateful to the anonymous referee for providing comments and suggestions that helped to improve the paper. We would like to express our gratitude to Hakim Luthfi Malasan, Hesti Retno Tri Wulandari, Aprilia, and Moedji Raharto for useful discussions and suggestions. Special thanks to Rhorhom Priyatikanto for sharing valuable insights and explanations for the improvements of this paper. M.B.N.K. acknowledges support from the National Natural Science Foundation of China (grant 11573004). This research was supported by the Research Development Fund (grant RDF$-$16$-$01$-$16) of Xi'an Jiaotong-Liverpool University (XJTLU). This work has been done using the facilities of Chalawan High-Performance Computing at the National Astronomical Research Institute of Thailand (NARIT) and supported by the computing facilities in the Department of Astronomy, Institut Teknologi Bandung.


\section*{Data Availability}
The data underlying this article will be shared on reasonable request to the corresponding author.

\bibliographystyle{mnras}
\bibliography{Darma_binary_clusters} 
\label{lastpage}
\end{document}